\documentclass[aps,prl,reprint,superscriptaddress,longbibliography]{revtex4-2}
\usepackage{amsmath,amssymb,bm,graphicx,booktabs}
\usepackage{fix-cm}
\usepackage[hidelinks]{hyperref}
\begin{document}
\title{Topological Pseudo-Goldstone Modes from Weakly Broken Dipole Conservation}
\author{Yan-Guang Yue}
\affiliation{Department of Physics and State Key Laboratory of Surface Physics, Fudan University, Shanghai 200433,  China}
\author{Jie Lou}
\affiliation{Department of Physics and State Key Laboratory of Surface Physics, Fudan University, Shanghai 200433,  China}
\author{Yan Chen}
\email{yanchen99@fudan.edu.cn}
\affiliation{Department of Physics and State Key Laboratory of Surface Physics, Fudan University, Shanghai 200433,  China}
\affiliation{Shanghai Branch, Hefei National Laboratory, Shanghai 201315,  China}
\date[]{}
\begin{abstract}
Weakly breaking a continuous symmetry in an ordered phase can convert its Goldstone modes into topological collective bands. We establish this mechanism in a neutral dipole condensate whose conserving parent supports two independent Goldstone phase modes. Weak real flavor mixing pins the phases and selects a chiral condensate. The resulting gyroscopic coupling acts on their winding polarization texture, producing a pseudo-Goldstone sector containing a topological lowest band. Restoring dipole conservation forces the two modes to meet at zero frequency, where their separate band Chern numbers cease to be defined. The same parent dipole charges define a tensor current whose circular response permits model-assisted reconstruction of low-energy embedded local Berry curvature in the benchmark regime. Analytical stability bounds and a large-occupation limit provide a controlled realization of this mechanism.
\end{abstract}
\maketitle

\textit{Introduction.---}Goldstone modes encode broken symmetries. Weak explicit breaking usually pins them; for coupled modes, it can also reshape their polarization and Berry geometry. Dipole conservation provides a setting for connecting symmetry-protected soft modes with band topology. Higher-moment symmetries constrain mobility~\cite{Pretko2017Subdimensional,Gromov2024Review} and allow unconventional superfluids~\cite{Yuan2020,Chen2021FractonicII}. In its conserving limit, a neutral dipole condensate spontaneously breaks two dipole rotations while preserving ordinary charge symmetry~\cite{Lake2022,Stahl2022Multipole,Jain2023,Armas2024Ideal}, giving two independent soft phases.

Topological dipolar phases~\cite{Fliss2021,MayMann2021,Lam2024} and topological Bogoliubov excitations~\cite{Furukawa2015,Bardyn2016,Xu2016,DiLiberto2016} establish two relevant precedents. Anakru and Bi showed that filled fermionic Chern bands generate a first-time-derivative coupling for dipolar Goldstone fields~\cite{Anakru2024}. Here the Chern numbers belong to the collective phase-mode bands themselves, and the gyroscopic coupling vanishes when dipole conservation is restored. Jalali-Mola \textit{et al.} obtained chiral condensates and Chern Bogoliubov bands from a frustrated kagome model with topologically trivial single-particle bands~\cite{Jalali2023}. How can a phase-mode sector protected by a conservation law itself acquire band topology under weak explicit breaking, and can the current associated with that law read the resulting geometry?

We establish a mechanism by which weak breaking of dipole conservation topologizes the phase-mode sector of a neutral condensate. Real flavor mixing pins the two phases and selects chiral order, generating a gyroscopic coupling that acts on their winding-two polarization. The resulting pseudo-Goldstone sector contains a topological lowest band. Restoring conservation forces the modes to meet at zero frequency, terminating their separate band invariants. The same parent dipole charges define a tensor current that reads the soft-mode geometry. Building on geometric tomography and circular spectroscopy~\cite{Flaschner2016,Asteria2019,Tesfaye2025}, this current excites the phase modes directly from the condensate vacuum.

\begin{figure}[t]
\centering
\includegraphics[width=\columnwidth]{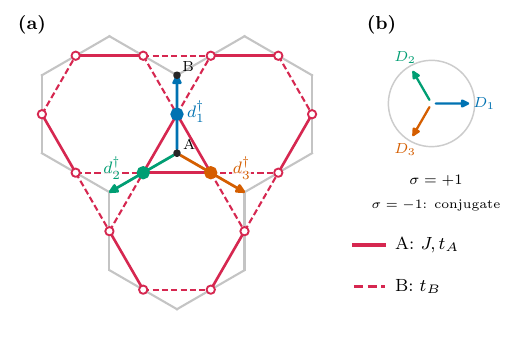}
\caption{Geometry and chiral order. (a) Three complete honeycomb hexagons (gray), with dipole orbitals at their bond centers. Arrows show the three central dipole vectors $\mathbf e_\nu=2\boldsymbol\delta_\nu$; their sum vanishes. Solid and dashed kagome links belong to $A$ and $B$ triangles, respectively. In the benchmark model, only $A$ triangles carry the three-body term $J$; flavor mixing acts on both. (b) The displayed uniform amplitudes have chirality $\sigma=+1$; complex conjugation gives $\sigma=-1$. The star term locks their common phase.}
\label{fig:lattice}
\end{figure}

\textit{Dipole-condensed parent.---}We work throughout at zero temperature. Under the higher-moment transformation $b_{\mathbf r}\mapsto e^{i(\alpha+\boldsymbol\beta\cdot\mathbf r)}b_{\mathbf r}$, a neutral bond dipole $b_{\mathbf r}^\dagger b_{\mathbf r+\mathbf e_\nu}$ transforms as $e^{i\boldsymbol\beta\cdot\mathbf e_\nu}$. Its phase therefore carries dipole charge while remaining neutral under the ordinary $U(1)$~\cite{Yuan2020,Lake2022}. We represent this sector by canonical soft-core fields $d_\alpha$, with $n_\alpha=d_\alpha^\dagger d_\alpha$, on honeycomb bond centers. This symmetry-based effective model is motivated by dipole order in constrained Bose--Hubbard systems~\cite{Lake2023Tilted,Zechmann2023}. The kagome geometry has three orientations $\nu$ per triangular cell. Their vectors sum to zero, allowing a cubic conversion that preserves vector dipole charge while locking the common phase. The parent Hamiltonian is
\begin{align}
H_0={}&(r+6t')\sum_\alpha n_\alpha
+u\sum_\alpha d_\alpha^{\dagger2}d_\alpha^2
\nonumber\\
&-t'\sum_{\langle\alpha\beta\rangle_\parallel}
(d_\alpha^\dagger d_\beta+\mathrm{H.c.})
+V\sum_{\langle\alpha\beta\rangle_K}n_\alpha n_\beta
\nonumber\\
&-J\sum_{\tau\in A}\left(\prod_{\alpha\in\tau}d_\alpha+\mathrm{H.c.}\right).
\label{eq:parent}
\end{align}
The same-flavor hopping connects six neighbors on each triangular sublattice; each kagome density-interaction bond is counted once. The three-body star term acts only on $A$ triangles (Fig.~\ref{fig:lattice}). The exact vector charge is
$\mathbf P_{\rm eff}=\sum_{\mathbf R,\nu}\mathbf e_\nu n_{\mathbf R\nu}$.
Equivalently, $N_1-N_3$ and $N_2-N_3$ are conserved, while the star term changes total dipole number. Here dipole condensation means $\langle d_\nu\rangle\ne0$: the two conserved vector charges protect its relative phases. Geometry and full conventions appear in the Supplemental Material (SM)~\cite{SM}.

We perturb the parent by real, positive flavor-mixing couplings,
\begin{equation}
H_{\rm mix}=\sum_{\tau\in A\cup B}t_\tau
\sum_{\alpha<\beta\in\tau}(d_\alpha^\dagger d_\beta+\mathrm{H.c.}).
\label{eq:mix}
\end{equation}
They explicitly break dipole conservation, while preserving time reversal and the common $Z_3$ rotation $d_\alpha\mapsto e^{2\pi i/3}d_\alpha$, which excludes a linear source. We use $t_A=0.8\lambda$, $t_B=0.2\lambda$, and $t_\Sigma=\lambda$ as a representative benchmark, and compare its topology with $t_A=t_B$ below. The remaining parameters are $\rho=1$, $u=2$, $V=1$, $J=0.5$. Stationarity fixes $r=t_\Sigma+J\rho-2(u+2V)\rho^2=\lambda-7.5$. Energies are in $t'=1$, triangular spacing is one, and $\hbar=1$. In this regime the chiral state is the global coherent-state minimum and its Gaussian Hessian is positive throughout the Brillouin zone (SM).

\textit{Weak breaking and chiral dynamics.---}For equal amplitudes $D_\nu=\rho e^{i\theta_\nu}$, the uniform phase potential is
\begin{equation}
f_\theta=-2J\rho^3\cos\Theta
+2t_\Sigma\rho^2\sum_{\mu<\nu}\cos(\theta_\mu-\theta_\nu),
\label{eq:phase}
\end{equation}
where $\Theta=\sum_\nu\theta_\nu$. At $\lambda=0$, $\Theta=0$ leaves a $U(1)^2$ orbit. The two commuting broken generators produce two phase Goldstone modes~\cite{WM2012}. At $t_\Sigma>0$, minimizing $|\sum_\nu e^{i\theta_\nu}|^2$ selects $120^\circ$ relative phases. The six labeled vacua are
\begin{equation}
D^{(s,\sigma)}=\rho e^{2\pi is/3}
(1,e^{i\sigma2\pi/3},e^{-i\sigma2\pi/3}),
\label{eq:vacua}
\end{equation}
with $s=0,1,2$ and chirality $\sigma=\pm1$.

\begin{figure}[t]
\centering
\includegraphics[width=\columnwidth]{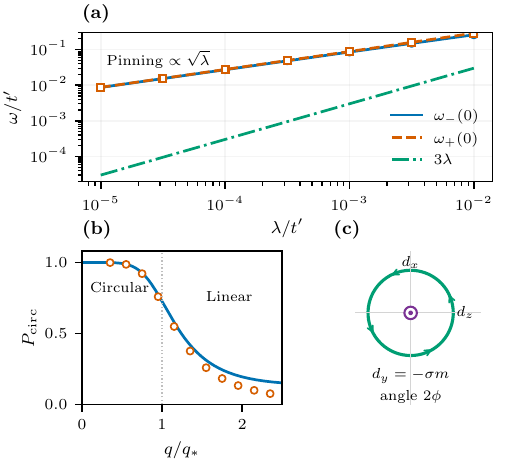}
\caption{From weak breaking to chiral dynamics in the benchmark completion. (a) Exact Gaussian poles, Eq.~\eqref{eq:gamma}, and full BdG symbols: pinning scales as $\sqrt\lambda$, while the splitting is $3\lambda$. (b) Circular polarization of the lower phase mode at $\lambda=10^{-4}$ along $\mathbf q=q\hat x$: $P_{\rm circ}=2\mathrm{Im}(p_x^*p_y)/|\mathbf p|^2$, with $\mathbf p=E^T(u_1-v_1)$ in the cell basis. The full BdG result (line) approaches $[1+(q/q_*)^6]^{-1/2}$ (circles) in the matching regime. (c) The real coefficients $(d_z,d_x)$ in Eq.~\eqref{eq:two} wind through $2\phi$; the gyroscopic mass $d_y=-\sigma m$ is perpendicular to this plane. All panels use $\sigma=+1$.}
\label{fig:gamma}
\end{figure}

The phase stiffness alone does not determine the mode polarizations. Their dynamics also retain the coupling to density fluctuations. Write $\theta_\nu=\theta_\nu^{(0)}+E_{\nu a}\varphi_a$, where $E_{\nu a}=\sqrt2 e_{\nu a}$ and $E^TE=I_2$. Eliminating the canonical amplitude variables gives, at leading order in gradients and weak breaking,
\begin{align}
\mathcal L_{\rm ph}={}&\frac{\chi_\lambda}{2}\dot{\boldsymbol\varphi}^{\,2}
+\sigma\chi_\lambda m(\varphi_x\dot\varphi_y-\varphi_y\dot\varphi_x)
\nonumber\\
&-\frac{\chi_\lambda\mu^2}{2}\boldsymbol\varphi^2
-\frac{\kappa}{2}(\partial_i\varphi_a)^2+\cdots,
\label{eq:eft}
\end{align}
where $a_v=J\rho+2(u-V)\rho^2$, $m=3t_\Sigma/2$, $\chi_\lambda=2\rho^2/(2a_v+m)$, $\mu^2=2a_vm$, and $\kappa=3t'\rho^2$. The omitted terms include mixing-induced gradient corrections. The first-time-derivative term rotates the two phase components into each other. Its sign is selected spontaneously by the condensate chirality, although the Hamiltonian is real. It vanishes with $\lambda$. At $\Gamma$ it gives the exact Gaussian poles:
\begin{equation}
\omega_\pm(0)=\sqrt{\mu^2+m^2}\pm m.
\label{eq:gamma}
\end{equation}
Phase pinning therefore scales as $\sqrt\lambda$, whereas the circular-mode splitting is exactly $3t_\Sigma\propto\lambda$ (Fig.~\ref{fig:gamma}). The restoring force and the handedness of the motion have distinct soft scales, even though they originate in the same weak perturbation.

\begin{figure}[t]
\centering
\includegraphics[width=\columnwidth]{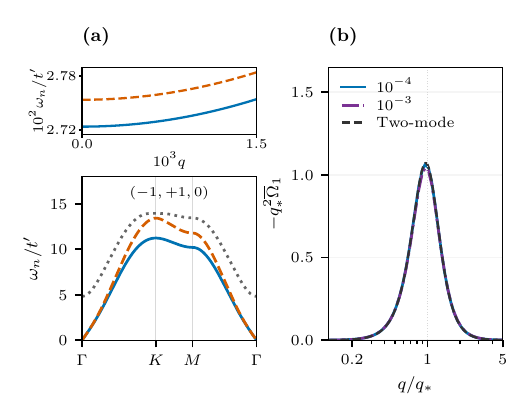}
\caption{Lattice topology and local geometry of the benchmark completion. (a) Full positive-frequency bands at $\lambda=10^{-4}$, $\sigma=+1$, with lowest-band Chern number $C_1=-1$ and full vector $(-1,+1,0)$. Upper zoom: absolute frequencies of the two low bands near $\Gamma$ along $q\hat x$, resolving their splitting $3\lambda$. (b) Angular-averaged embedded BdG curvature at two mixing strengths and the two-mode form, Eq.~\eqref{eq:berry}. Curvature peaks near the polarization crossover $q_*=(m/|a|)^{1/3}$.}
\label{fig:topology}
\end{figure}

\textit{Topological pseudo-Goldstone sector.---}Gapping alone is not topological; the Berry curvature arises because the gyroscopic mass gaps a momentum-space polarization winding. In the benchmark conserving parent, the two phase modes have the same leading gradient stiffness and sound velocity. Eliminating the gapped common-phase sector leaves a longitudinal correction proportional to $\mathbf q\mathbf q^T$ in the amplitude kernel (SM). With the $q^2$ phase stiffness, this splits squared frequencies at $q^4$, giving $\omega_L-\omega_T=2aq^3$. The longitudinal and transverse polarization projectors rotate through $2\phi$, where $\phi$ is the momentum angle. Projection onto the positive-frequency doublet gives
\begin{equation}
 h_{\rm low}-\bar\omega I_2=
 f(q)[\cos2\phi\,\sigma_z+\sin2\phi\,\sigma_x]
 -\sigma m\sigma_y,
\label{eq:two}
\end{equation}
where the benchmark has $f(q)=aq^3$ with $a\simeq0.084$, and the Pauli matrices act on the relative-phase doublet. The first term favors linear polarization along the momentum-dependent axes. The gyroscopic term favors circular motion and supplies a mass perpendicular to the winding plane. Their competition gives nearly circular modes for $|f|\ll m$ and nearly linear modes for $|f|\gg m$, as shown in Fig.~\ref{fig:gamma}(b,c). Reversing the condensate chirality reverses the mass.

The benchmark low-band curvature is
\begin{equation}
\Omega_1(q)\simeq-
\sigma\frac{3ma^2q^4}{(m^2+a^2q^6)^{3/2}},
\qquad q_*=(m/|a|)^{1/3}.
\label{eq:berry}
\end{equation}
It is concentrated near the polarization crossover $|f(q_*)|=m$ [Fig.~\ref{fig:topology}(b)]. The benchmark therefore has three distinct scales: phase pinning $\propto\lambda^{1/2}$, chiral splitting $\propto\lambda$, and a curvature radius $q_*\propto\lambda^{1/3}$. The projection is controlled in the Gaussian matching regime $q_{\rm pg}\sim\lambda^{1/2}\ll q\sim q_*$, where the positive doublet separates from its negative-frequency partners. The winding structure persists in the star controls below, whereas $f(q)\sim q^3$ and $q_*\sim\lambda^{1/3}$ depend on the benchmark gradient structure. With both stars present, unequal parent sound velocities instead give a leading linear splitting (SM).

We establish the global topology with the full six-component lattice problem, $\Sigma_z\mathcal H_{\rm BdG}w_n=\omega_nw_n$, where $\Sigma_z=\operatorname{diag}(I_3,-I_3)$ and $w_n^\dagger\Sigma_zw_m=\delta_{nm}$ in the positive sector~\cite{Shindou2013,Tesfaye2025}. With $\mathcal A_n=i w_n^\dagger\Sigma_z\nabla_{\mathbf k}w_n$, the benchmark completion at $\lambda=10^{-4},10^{-3}$ gives
\begin{equation}
(C_1,C_2,C_3)=(-1,+1,0)\qquad(\sigma=+1),
\label{eq:chern}
\end{equation}
with sign reversal for the opposite chirality [Fig.~\ref{fig:topology}(a)]. Mesh refinement and the periodic and properly sewn embedded bases agree~\cite{FHS2005,SM}. The same Chern vector persists for $t_A=t_B$ with the original star term.

To separate the soft-mode mechanism from the lattice allocation of Chern charge, we redistribute the star conversion at fixed $J_A+J_B=J$. Numerically, the lowest band remains isolated with $C_1=-1$ throughout the tested paths, including at the fully symmetric endpoint $J_A=J_B$ and $t_A=t_B$. The Chern vector changes to $(-1,0,+1)$ through a finite-frequency touching of bands 2 and 3 at a Brillouin-zone corner, before the equal-star endpoint. This transfer of compensating Chern charge takes place with a positive Gaussian Hessian (SM).

The winding--mass doublet gives a local Berry-flux contribution approaching $-\sigma$ in a benchmark disk $q_*\ll R\ll1$. The full Brillouin zone fixes the absolute Chern numbers and how compensating charge is distributed among higher bands. At $\lambda=0$, exact dipole conservation forces both Goldstone frequencies to vanish at $\Gamma$, and separate band Chern numbers cease to be defined. Symmetry restoration thus fixes the zero-frequency endpoint of the topological pseudo-Goldstone sector.

\begin{figure}[t]
\centering
\includegraphics[width=\columnwidth]{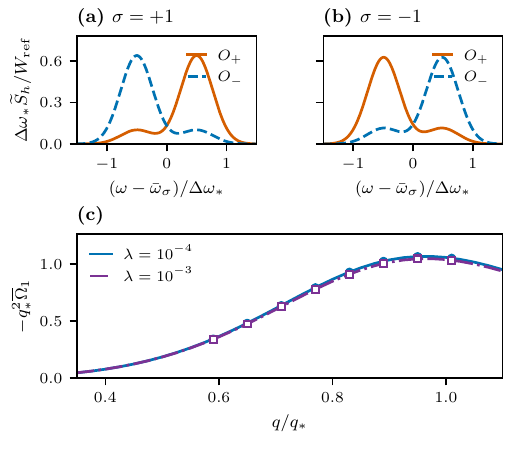}
\caption{Tensor-current spectroscopy of the benchmark phase doublet. (a,b) Independently evaluated spectra for opposite chiralities at the same displayed momentum $\mathbf q=q_*\hat x$, $\lambda=10^{-4}$. The helicity preferences reverse; exact time reversal relates $(\sigma,\mathbf q)$ to $(-\sigma,-\mathbf q)$. (c) Independent full BdG curvature (lines) and reconstruction from jointly fitted helicity areas (symbols), using noiseless spectra with sharp intrinsic poles. The spectral curves in (a,b) are Gaussian-convolved with standard deviations $(s_\omega/\Delta\omega_*,s_q/q_{\rm res})=(0.25,0.1)$ and one shared intensity calibration $W_{\rm ref}$. Here $\Delta\omega_*$ is the splitting at $q_*\hat x$ and $q_{\rm res}=\Delta\omega_*/|\partial_q\bar\omega|$ (SM); each spectrum is centered on its own $\bar\omega_\sigma$. The SM gives the complete finite-resolution and finite-count tests, including unsuccessful reconstructions.}
\label{fig:probe}
\end{figure}

\textit{Tensor spectroscopy.---}Because the soft coordinates are dipole phases, the parent charges define a tensor current that couples directly to their polarization texture. At finite mixing its continuity equation acquires a symmetry-breaking torque (SM). A Peierls field $a_{ia}$ on the same-flavor links defines $J_i^a=-\delta H/\delta a_{ia}$, with $i$ the transport direction and $a=x,y$ the dipole component. Using the common-cell source convention specified in the SM, its one-quasiparticle vertex couples to the phase amplitudes: $J_i^{a(1)}\simeq i\kappa q_i\varphi_a$. The symmetric traceless circular combinations are
\begin{equation}
 O_\pm=\tfrac12[J_x^x-J_y^y\pm i(J_x^y+J_y^x)].
\label{eq:probe}
\end{equation}
They preferentially excite opposite members of the doublet from the condensate vacuum. The independently calculated spectra at fixed $\mathbf q$ show opposite helicity preferences [Fig.~\ref{fig:probe}(a,b)]. Exact time reversal relates $(\sigma,\mathbf q)$ to $(-\sigma,-\mathbf q)$, so it does not impose equality of the fixed-momentum spectra after a helicity exchange. Because the linear current vertex vanishes at $q=0$, the measurement requires finite momentum.

Let $W_{\pm,n}=|\langle0|O_\pm|n,\mathbf q\rangle|^2$ and $\mathcal D_1(q,\phi)=(W_{+,1}-W_{-,1})/(W_{+,1}+W_{-,1})$. Extract both areas with a common source calibration, form the contrast at each angle, then average: $\overline{\mathcal D}_1(q)=(2\pi)^{-1}\int_0^{2\pi}d\phi\,\mathcal D_1(q,\phi)$. The angular average of the embedded curvature is $\overline{\Omega}_1(q)=(2\pi)^{-1}\int_0^{2\pi}d\phi\,\Omega_1(q,\phi)$. In the benchmark low-energy matching regime,
\begin{align}
\overline{\mathcal D}_1(q)&\simeq
-\frac{\sigma m}{\sqrt{m^2+f(q)^2}},\nonumber\\
\overline{\Omega}_1(q)&\simeq
-\frac1q\frac{d\overline{\mathcal D}_1(q)}{dq}.
\label{eq:tomography}
\end{align}
Thus momentum-resolved dichroism reconstructs the embedded local Berry curvature within the low-energy matching regime (Fig.~\ref{fig:probe}). At $q=q_*$, the relative differences from independent full BdG curvature are $0.033\%$ and $0.135\%$ for $\lambda=10^{-4},10^{-3}$. The derivative measures how the polarization tilts away from circular motion as the winding term overtakes the mass. The result is a model-assisted local reconstruction~\cite{Tran2017,Tesfaye2025}; high-band admixture, lattice anisotropy, and orbital embedding give corrections.

Finite-resolution simulations convolve the full spectra with Gaussian frequency and two-dimensional momentum kernels, then jointly fit both helicities with one intensity calibration. In the resolved regime, the noiseless reconstruction agrees with the independent BdG curvature [Fig.~\ref{fig:probe}(c)]. The required momentum resolution becomes increasingly stringent near conservation; broader kernels can produce large finite-count errors even when fits converge. The SM quantifies the resolution scales, count assumptions, and unsuccessful reconstructions.

\textit{Stability.---}An analytical energy bound establishes the chiral state as the global coherent-state minimum in a finite parameter regime and gives a positive, momentum-independent lower bound on $\mathcal H_{\rm BdG}$. A large-occupation family leaves the Gaussian spectrum unchanged while suppressing nonlinear corrections at fixed $\lambda>0$. At both benchmarks, the modes lie below the leading zero-temperature two-quasiparticle decay threshold at sampled momenta in the spectroscopic window of Fig.~\ref{fig:probe}~\cite{Beliaev2024,SM}. These threshold tests constrain leading decay; higher-order lifetimes remain open.

\textit{Discussion.---}Three ingredients organize the route identified here: a symmetry-protected multicomponent soft sector, momentum-dependent polarization winding, and a transverse gyroscopic mass generated by weak breaking. Symmetry fixes the zero-frequency Goldstone endpoint; the winding--mass dynamics~\cite{Sun2009} generate local Berry curvature, while the full lattice determines the band Chern numbers. The star interpolation demonstrates that lowest-band topology can persist through a change in the higher-band Chern allocation. Equations~\eqref{eq:parent} and~\eqref{eq:mix} provide a minimal symmetry-based realization in the neutral dipole sector.

Strong indirect bulk-band overlap motivates the bulk tensor probe: it leaves no common frequency gap protecting an isolated boundary channel~\cite{Malki2019,SM}. In the benchmark, the Gaussian resolution scale $q_{\rm res}\sim\lambda$ shrinks faster than the geometric scale $q_*\sim\lambda^{1/3}$: spectral access becomes more demanding as the low-energy description improves. Proposals for synthetic tensor fields and dipole-current measurements~\cite{Zhang2025Tensor,Zhang2026} motivate implementation of the probe; a microscopic realization of the star conversion remains an open direction. The tensor response reveals how the parent conservation law organizes the geometry of the collective excitations produced by its weak breaking.

\noindent\begin{minipage}{\columnwidth}
\section*{ACKNOWLEDGMENTS}
We thank Fei Teng for his helpful discussions. This work is supported by the National Key Research and Development Program of China (Grant No. 2022YFA1404204), the National Natural Science Foundation of China (Grant No. 12274086), and the Quantum Science and Technology-National Science and Technology Major Project (Grant No. 2024ZD0300104).
\end{minipage}

\makeatletter
\immediate\write\@auxout{\string\citation{ArticleTitleControl}}
\makeatother
\nocite{Pretko2018Gauge,Gromov2019Multipole,Stahl2023Hydro,Colpa1978,Peano2018,Engelhardt2015,Ozawa2018,Xiao2010,Price2012}
\bibliographystyle{apsrev4-2}
\bibliography{references}
\end{document}


\title{Supplemental Material for\\Topological Pseudo-Goldstone Modes from Weakly Broken Dipole Conservation}
\author{Yan-Guang Yue}
\affiliation{Department of Physics and State Key Laboratory of Surface Physics, Fudan University, Shanghai 200433,  China}
\author{Jie Lou}
\affiliation{Department of Physics and State Key Laboratory of Surface Physics, Fudan University, Shanghai 200433,  China}
\author{Yan Chen}
\email{yanchen99@fudan.edu.cn}
\affiliation{Department of Physics and State Key Laboratory of Surface Physics, Fudan University, Shanghai 200433,  China}
\affiliation{Shanghai Branch, Hefei National Laboratory, Shanghai 201315,  China}
\date[]{}
\maketitle
\renewcommand{\theequation}{S\arabic{equation}}
\setcounter{equation}{0}
\renewcommand{\thefigure}{S\arabic{figure}}
\renewcommand{\thetable}{S\Roman{table}}

\noindent This supplement gives the effective lattice model, derivations, and numerical methods behind the Letter. The zero-temperature analysis covers the chiral saddle, Gaussian stability and topology, tensor spectroscopy, finite-resolution inference, and sampled decay thresholds. The canonical fields provide a symmetry-based description of the neutral dipole sector.

\section{Lattice model, dipole representation, and chiral saddle}
\subsection{Geometry and bond counting}
The triangular Bravais vectors, reciprocal vectors, and cell area are
\begin{align}
\mathbf a_1&=(1,0),&\mathbf a_2&=(1/2,\sqrt3/2),&A_c&=\sqrt3/2,\\
\mathbf b_1&=2\pi(1,-1/\sqrt3),&\mathbf b_2&=2\pi(0,2/\sqrt3).
\end{align}
The three dipole orbitals sit at
\begin{equation}
\boldsymbol\delta_1=(0,1/(2\sqrt3)),\quad
\boldsymbol\delta_2=(-1/4,-1/(4\sqrt3)),\quad
\boldsymbol\delta_3=(1/4,-1/(4\sqrt3)),\qquad
\mathbf e_\nu=2\boldsymbol\delta_\nu.
\end{equation}
An $A$ triangle contains $(\mathbf R,1),(\mathbf R,2),(\mathbf R,3)$; a $B$ triangle contains $(\mathbf R,1),(\mathbf R+\mathbf a_2,2),(\mathbf R+\mathbf a_2-\mathbf a_1,3)$. Each kagome edge is counted once. Parallel-bond hopping means hopping between the same dipole orientation, with displacements $\pm\mathbf a_1$, $\pm\mathbf a_2$, and $\pm(\mathbf a_1-\mathbf a_2)$. It is not restricted to motion along $\mathbf e_\nu$.

With $[d_\alpha,d_\beta^\dagger]=\delta_{\alpha\beta}$, the Hamiltonian is
\begin{align}
H={}&(r+6t')\sum_\alpha n_\alpha+u\sum_\alpha d_\alpha^{\dagger2}d_\alpha^2
-t'\sum_{\langle\alpha\beta\rangle_\parallel}(d_\alpha^\dagger d_\beta+\mathrm{H.c.})
+V\sum_{\langle\alpha\beta\rangle_K}n_\alpha n_\beta\nonumber\\
&-J\sum_{\tau\in A}\left(\prod_{\alpha\in\tau}d_\alpha+\mathrm{H.c.}\right)
+\sum_{\tau\in A\cup B}t_\tau\sum_{\alpha<\beta\in\tau}(d_\alpha^\dagger d_\beta+\mathrm{H.c.}).
\label{eq:Smodel}
\end{align}
The on-site term is normal ordered. Its coherent-state expectation is $u|D_\alpha|^4$. In particular, $V$ denotes the interaction \emph{per edge}: the uniform cross-amplitude coefficient is $v_L=2V$. All couplings are real. We use $t_A=0.8\lambda$, $t_B=0.2\lambda$, $t_\Sigma=t_A+t_B=\lambda$ and, unless stated otherwise,
\begin{equation}
\rho=1,\qquad t'=1,\qquad u=2,\qquad V=1,\qquad J=0.5.
\end{equation}

\subsection{Symmetry representation and neutral phase fields}
At $\lambda=0$, the two commuting charges $N_1-N_3$ and $N_2-N_3$ are conserved, or equivalently $\mathbf P_{\rm eff}=\sum_{\mathbf R\nu}\mathbf e_\nu n_{\mathbf R\nu}$. The star conversion changes all three occupations by the same amount. Its dipole charge vanishes because $\sum_\nu\mathbf e_\nu=0$.

The transformation of a neutral microscopic representative, $\mathcal O_\nu=b_{\mathbf r}^\dagger b_{\mathbf r+\mathbf e_\nu}$, is $\mathcal O_\nu\mapsto e^{i\boldsymbol\beta\cdot\mathbf e_\nu}\mathcal O_\nu$ under $b_{\mathbf r}\mapsto e^{i(\alpha+\boldsymbol\beta\cdot\mathbf r)}b_{\mathbf r}$. This matches the effective dipole representation, but does not establish a canonical operator projection. Charge and dipole condensation have distinct low-energy descriptions~\cite{Yuan2020,Lake2022,Jain2023}. In the present neutral sector the two phases are independent; identifying them with the gradient of one scalar phase would remove the transverse mode.

Let
\begin{equation}
E_{\nu a}=\sqrt8\delta_{\nu a}=\sqrt2e_{\nu a},\quad
E^TE=I_2,\quad E^T\mathbf s=0,\quad \mathbf s=(1,1,1)^T.
\end{equation}
The normalized charges are $Q_a=\sum_{\mathbf R\nu}E_{\nu a}n_{\mathbf R\nu}=\sqrt2P_{{\rm eff},a}$. The fields $\theta_\nu=\theta_\nu^{(0)}+E_{\nu a}\varphi_a$ inherit two independent constant shifts. At finite mixing,
\begin{equation}
[\mathbf P_{\rm eff},d_\mu^\dagger d_\nu]
=(\mathbf e_\mu-\mathbf e_\nu)d_\mu^\dagger d_\nu,
\end{equation}
so the full model is not exactly dipole conserving. It retains the common $Z_3$ rotation and time reversal before a vacuum is chosen.

\subsection{Phase selection, amplitude condition, and saddle}
The equal-amplitude phase potential is
\begin{equation}
f_\theta=-2J\rho^3\cos\Theta+2t_\Sigma\rho^2
\sum_{\mu<\nu}\cos(\theta_\mu-\theta_\nu),\qquad\Theta=\sum_\nu\theta_\nu.
\end{equation}
Since $|\sum_\nu e^{i\theta_\nu}|^2\ge0$, the mixing term is bounded below by $-3t_\Sigma\rho^2$. Equality requires the three equal-length complex vectors to sum to zero, hence their relative angles are $120^\circ$. Combining this with $\Theta=0$ gives six labeled vacua,
\begin{equation}
D^{(s,\sigma)}=\rho e^{2\pi is/3}(1,e^{i\sigma2\pi/3},e^{-i\sigma2\pi/3}),
\qquad s=0,1,2,\quad\sigma=\pm1.
\end{equation}
They have two chiralities, not only two field configurations. At $\lambda=0$, the relative phases belong to a continuous global symmetry orbit and cannot be selected by symmetry-preserving zero-point corrections.

For arbitrary uniform amplitudes in the parent, fixing $S=\sum_\nu\rho_\nu^2$ gives
\begin{equation}
u\sum_\nu\rho_\nu^4+v_L\sum_{\mu<\nu}\rho_\mu^2\rho_\nu^2
=\frac{u+v_L}{3}S^2+\frac{2u-v_L}{6}
\sum_{\mu<\nu}(\rho_\mu^2-\rho_\nu^2)^2.
\end{equation}
Together with $\rho_1\rho_2\rho_3\le(S/3)^{3/2}$, this gives the uniform equal-amplitude minimum at each $S$ when $v_L\le2u$ and $J>0$. The stronger all-spatial statement used in the Letter is proved below. For the selected branch,
\begin{equation}
f_{\rm ch}=3(r-t_\Sigma)\rho^2-2J\rho^3+3(u+2V)\rho^4,
\end{equation}
so stationarity requires
\begin{equation}
r=t_\Sigma+J\rho-2(u+2V)\rho^2,
\qquad M_0=J\rho+2u\rho^2+t_\Sigma.
\label{eq:Ssaddle}
\end{equation}
The constant-amplitude benchmark is thus a slice $r=\lambda-7.5$, not a scan at fixed bare mass.

\section{All-spatial coherent-state bound and Gaussian positivity}
For arbitrary spatial coherent fields $z_\alpha$, define $x_\alpha=|z_\alpha|$ and $S_\tau=\sum_{\alpha\in\tau}x_\alpha^2$. Exact bond counting yields
\begin{align}
u\sum_\alpha x_\alpha^4+V\sum_{\langle\alpha\beta\rangle_K}x_\alpha^2x_\beta^2
&=(u-V)\sum_\alpha x_\alpha^4+\frac V2\sum_\tau S_\tau^2,\\
-2t'\sum_\parallel\mathrm{Re}(z_\alpha^*z_\beta)
&=-6t'\sum_\alpha x_\alpha^2+t'\sum_\parallel|z_\alpha-z_\beta|^2,\\
2t_\tau\sum_{\alpha<\beta\in\tau}\mathrm{Re}(z_\alpha^*z_\beta)
&=t_\tau\left|\sum_{\alpha\in\tau}z_\alpha\right|^2-t_\tau S_\tau.
\end{align}
Using Eq.~\eqref{eq:Ssaddle}, the difference from the uniform chiral state is
\begin{align}
E[z]-E_0={}&t'\sum_\parallel|z_\alpha-z_\beta|^2+
\sum_\tau t_\tau\left|\sum_{\alpha\in\tau}z_\alpha\right|^2
+\frac V2\sum_\tau(S_\tau-3\rho^2)^2+\sum_\alpha R(x_\alpha)\nonumber\\
&+\frac{2J}{3}\sum_{\tau\in A}\left[
\sum_{\alpha\in\tau}x_\alpha^3-3\mathrm{Re}\prod_{\alpha\in\tau}z_\alpha\right],\label{eq:Sbound}\\
R(x)={}&(x-\rho)^2\left[(u-V)(x+\rho)^2-\frac{2J}{3}\left(x+\frac\rho2\right)\right].
\end{align}
The final line of the energy difference is nonnegative by the arithmetic--geometric mean inequality. If
\begin{equation}
u-V\ge\frac{J}{3\rho},\qquad V\ge0,\quad J,t'>0,\quad t_A,t_B\ge0,
\label{eq:Scondition}
\end{equation}
then $R(x)\ge0$ for all $x\ge0$. All terms in Eq.~\eqref{eq:Sbound} vanish on the uniform chiral vacuum, proving global minimality within spatial coherent states. The condition is sufficient, not a full phase boundary. At the benchmark, $u-V=1>1/6$ and $R(x)\ge(5/6)(x-1)^2$.

For $t_A>0$, a subset of these forms already establishes a strict quadratic bound. Set $z_\nu=e^{i\theta_\nu}[\rho+(Q_\nu+iP_\nu)/\sqrt2]$, $P_\perp=I_3-\mathbf s\mathbf s^T/3$, $\chi_\nu=e^{i\theta_\nu}$, and $W=\chi^*\chi^T$. With $j=J\rho$ and $R_0=4(u-V)\rho^2-j$, the retained on-cell matrix is
\begin{equation}
K_{\rm loc}=\begin{pmatrix}X_{\rm loc}&C_{\rm loc}\\C_{\rm loc}^T&Y_{\rm loc}\end{pmatrix},
\end{equation}
\begin{equation}
X_{\rm loc}=R_0I_3+3jP_\perp+t_A\mathrm{Re}W,\quad
Y_{\rm loc}=j\mathbf s\mathbf s^T+t_A\mathrm{Re}W,\quad
C_{\rm loc}=-t_A\mathrm{Im}W.
\end{equation}
Its eigenvalues are $R_0$, $3j$, and $a_v+m_A\pm\sqrt{a_v^2+m_A^2}$, the last pair each twice degenerate, where $m_A=3t_A/2$. Consequently
\begin{equation}
\varepsilon=\min\left\{R_0,3j,
\frac{2a_vm_A}{a_v+m_A+\sqrt{a_v^2+m_A^2}}\right\}>0.
\end{equation}
The discarded forms are positive semidefinite. A Fourier transform and the unitary conversion from $Q,P$ to Nambu coordinates preserve the bound, hence
\begin{equation}
\mathcal H_{\rm BdG}(\mathbf k)\succeq\varepsilon I_6
\quad\text{for every }\mathbf k.
\end{equation}
At $\lambda=10^{-4},10^{-3}$ the bound is $1.1999712\times10^{-4}$ and $1.1997120\times10^{-3}$, respectively. This proves quadratic energetic and dynamical stability under the assumptions, but neither an interband gap nor the exact interacting ground state. It tends to zero as $t_A\to0$.

\section{Quadratic expansion, Fourier conventions, and Chern computation}
\subsection{Real-space expansion and Bloch blocks}
Write $d_\alpha=e^{i\theta_\alpha}(\rho+a_\alpha)$. The density is
\begin{equation}
n_\alpha=\rho^2+\rho(a_\alpha+a_\alpha^\dagger)+a_\alpha^\dagger a_\alpha.
\end{equation}
The phase-blind density interaction generates identical off-diagonal normal and anomalous coefficients at quadratic order. Its normal contribution must not acquire an extra chiral phase. The star contributes
\begin{equation}
H_J^{(2)}=-J\rho\sum_{\tau\in A}
(a_1a_2+a_2a_3+a_3a_1+\mathrm{H.c.}),
\end{equation}
which is anomalous in the rotated basis. Its real-space coefficients are real, although its embedded momentum-space matrix need not be real.

For $a_{\mathbf R\nu}=N_c^{-1/2}\sum_{\mathbf k}e^{i\mathbf k\cdot\mathbf R}a_{\mathbf k\nu}$, define $z_{\mu\nu}=e^{2i\mathbf k\cdot(\boldsymbol\delta_\mu-\boldsymbol\delta_\nu)}$. The complete blocks are
\begin{align}
A_{\nu\nu}&=M_0+\gamma, & B_{\nu\nu}&=2u\rho^2,\\
A_{\mu\nu}&=V\rho^2(1+z_{\mu\nu})+(t_A+t_Bz_{\mu\nu})e^{i(\theta_\nu-\theta_\mu)},
&B_{\mu\nu}&=V\rho^2(1+z_{\mu\nu})-J\rho\quad(\mu\ne\nu),\\
\gamma(\mathbf k)&=2t'[3-\cos(\mathbf k\cdot\mathbf a_1)-\cos(\mathbf k\cdot\mathbf a_2)-\cos(\mathbf k\cdot(\mathbf a_1-\mathbf a_2))].&&
\end{align}
In Nambu coordinates $\Psi_{\mathbf k}=(a_{\mathbf k},a_{-\mathbf k}^\dagger)^T$,
\begin{equation}
H^{(2)}=\frac12\sum_{\mathbf k}\Psi_{\mathbf k}^\dagger
\begin{pmatrix}A(\mathbf k)&B(\mathbf k)\\B^\dagger(\mathbf k)&A^T(-\mathbf k)\end{pmatrix}
\Psi_{\mathbf k}+\text{constant}.
\end{equation}
The checks are $A=A^\dagger$, $B(\mathbf k)=B^T(-\mathbf k)$, and Nambu redundancy. For Hermitian $A$, $A^T(-\mathbf k)=A^*(-\mathbf k)$; neither is generally $A^\dagger(-\mathbf k)$.

The embedded coefficients are obtained with $U(\mathbf k)=\operatorname{diag}(e^{-i\mathbf k\cdot\boldsymbol\delta_\nu})$ and $\mathcal U=\operatorname{diag}[U(\mathbf k),U^*(-\mathbf k)]$. Thus $\mathcal H_{\rm emb}=\mathcal U\mathcal H_{\rm cell}\mathcal U^\dagger$. Explicitly,
\begin{align}
A_{\mu\nu}^{\rm emb}&=2V\rho^2\cos(\mathbf k\cdot\Delta_{\mu\nu})+
[t_Ae^{-i\mathbf k\cdot\Delta_{\mu\nu}}+t_Be^{i\mathbf k\cdot\Delta_{\mu\nu}}]e^{i(\theta_\nu-\theta_\mu)},\\
B_{\mu\nu}^{\rm emb}&=2V\rho^2\cos(\mathbf k\cdot\Delta_{\mu\nu})-J\rho e^{-i\mathbf k\cdot\Delta_{\mu\nu}},
\qquad \Delta_{\mu\nu}=\delta_\mu-\delta_\nu.
\end{align}
The anomalous phase follows the creation-pair block and the stated Fourier convention. At a reciprocal boundary, embedded eigenvectors must be sewn by $\mathcal U(\mathbf G)$, rather than copied periodically. Consistent changes of representation do not create a Chern number.

\subsection{Stable modes and discretized invariants}
For a positive-definite $\mathcal H=R^\dagger R$, diagonalizing the Hermitian matrix $R\Sigma_zR^\dagger$ gives a stable route to the spectrum. A positive eigenvector $y_n$ of eigenvalue $\omega_n$ gives $w_n=R^{-1}y_n\sqrt{\omega_n}$, normalized by $w_n^\dagger\Sigma_zw_m=\delta_{nm}$. Indefinite or zero-mode cases are not assigned $C=0$ by this routine. They require separate treatment~\cite{Shindou2013,Tesfaye2025}.

We take $\mathcal A_n=i w_n^\dagger\Sigma_z\nabla w_n$. On a positively oriented reciprocal mesh, set
\begin{equation}
L_\alpha(\mathbf k)=
\frac{w_n^\dagger(\mathbf k)\Sigma_z w_n(\mathbf k+\Delta\mathbf k_\alpha)}
{|w_n^\dagger(\mathbf k)\Sigma_z w_n(\mathbf k+\Delta\mathbf k_\alpha)|}.
\end{equation}
For this overlap orientation and the $+i$ connection convention, the FHS sum is
\begin{equation}
C_n=-\frac1{2\pi}\sum_{\mathbf k}\operatorname{Arg}
[L_1(\mathbf k)L_2(\mathbf k+\Delta\mathbf k_1)
L_1(\mathbf k+\Delta\mathbf k_2)^{-1}L_2(\mathbf k)^{-1}].
\end{equation}
Near-zero overlaps require mesh refinement, not a regularizer hiding their phases. An integer on a finite mesh is not alone a convergence test~\cite{FHS2005}. Positive-band isolation, eigenpair residuals, and basis sewing are checked independently.

\begin{table}[t]
\caption{Full-BZ benchmarks. The minima are sampled values; the analytic energy-matrix bound addresses a different question. Both $72^2$ and $144^2$ meshes give the listed Chern vector.}
\begin{ruledtabular}
\begin{tabular}{ccccc}
$\lambda$&$\min\omega_1$&$\Delta_{12}^{\rm dir}$&$\Delta_{23}^{\rm dir}$&$(C_1,C_2,C_3)$\\
$10^{-4}$&0.02723654&0.00030000&0.52070352&$(-1,+1,0)$\\
$10^{-3}$&0.08511553&0.00299961&0.51846720&$(-1,+1,0)$
\end{tabular}
\end{ruledtabular}
\end{table}
The reversed chirality gives $(+1,-1,0)$. The equal-hopping check $t_A=t_B=0.0005$ remains nontrivial. The complete positive-frequency band bundle has zero total Chern number in the positive-definite setting~\cite{Shindou2013}; the results satisfy this check. At the conserving point the low pair touches at zero frequency, so separate global single-band invariants are not defined. For a complex parent representative, bare time reversal is followed by the allowed $U(1)^2$ rotation $d_\nu\mapsto e^{2i\theta_\nu^{(0)}}d_\nu$. This combination fixes the chosen vacuum. In the rotated fluctuation basis, $\mathcal H_0(-\mathbf k)=\mathcal H_0(\mathbf k)^*$, allowing smooth local curvature odd in momentum without requiring it to vanish everywhere.

Global Chern invariants agree between the periodic and sewn embedded representations. Local curvature need not agree pointwise under their momentum-dependent transformation. All radial curvature plots and direct reconstruction comparisons in this manuscript use the embedded orbital representation, whereas the source remains defined at the stated common cell coordinates. Their low-energy relation includes the corresponding subleading orbital corrections.

\subsection{Stationary hopping and chirality controls}
The hopping ratio is not a source of explicit time-reversal breaking. We independently checked $t_A/t_\Sigma=0.8$ and $0.5$ at both $t_\Sigma=\lambda$ benchmarks, always resetting the stationary coefficient to $r=\lambda-7.5$. All three complex first derivatives of the uniform energy vanish to $1.59\times10^{-15}$ or better. Table~\ref{tab:Smechanism} gives the sampled direct gaps. Every case has $(C_1,C_2,C_3)=(-1,+1,0)$ for $\sigma=+1$ on both $72^2$ and $144^2$ meshes. A $72^2$ calculation at the opposite chirality gives $(+1,-1,0)$. These finite-mesh tests are separate from the analytical energetic-positivity bound.

\begin{table}[t]
\caption{Stationary full-lattice controls. Gaps are minima on the $144^2$ mesh. The last column is the ideal angular-averaged curvature reconstruction error at $q_*$.}\label{tab:Smechanism}
\begin{ruledtabular}
\begin{tabular}{ccccc}
$\lambda$ & $t_A/t_\Sigma$ & $\Delta_{12}^{\rm dir}$ & $\Delta_{23}^{\rm dir}$ & Relative error\\
$10^{-4}$ & $0.8$ & $0.0003000000$ & $0.52070352$ & $0.03280\%$\\
$10^{-4}$ & $0.5$ & $0.0003000000$ & $0.52079897$ & $0.02325\%$\\
$10^{-3}$ & $0.8$ & $0.0029996124$ & $0.51846720$ & $0.13548\%$\\
$10^{-3}$ & $0.5$ & $0.0029995580$ & $0.51942170$ & $0.08388\%$\\
\end{tabular}
\end{ruledtabular}
\end{table}

We also test the exact time-reversal relations
\begin{align}
\omega_n(\mathbf q,\sigma)&=\omega_n(-\mathbf q,-\sigma),\\
W_{\pm,n}(\mathbf q,\sigma)&=W_{\mp,n}(-\mathbf q,-\sigma),\\
\Omega_n(\mathbf q,\sigma)&=-\Omega_n(-\mathbf q,-\sigma).
\end{align}
For 24 directions at $q/q_*=0.35,1,1.1$, all three opposite-momentum residuals are zero at stored double precision in this implementation. At fixed momentum, however, the largest frequency differences over these samples are $5.54\times10^{-6}$ and $1.19\times10^{-4}$ for the unequal-hopping benchmarks. Equal hopping does not impose inversion symmetry on the full model, which retains the $A$-only star term and the stated source convention. Consequently the main-figure helicity exchange at fixed $\mathbf q$ is a preference, not an identity of spectra.

For the equal-hopping cases, angular averaging over 24 directions gives $\overline{\mathcal D}_1(q_*)=-0.70880360,-0.71369060$ and independent scaled curvatures $-q_*^2\overline{\Omega}_1=1.05572425,1.03791863$. Doubling the angular resolution and halving the radial difference step changes the reconstructed scaled curvature by at most $1.81\times10^{-6}$ across all four parameter sets. 

Removing the mass from the reduced two-mode Hamiltonian makes its eigenvectors locally real away from the origin, where the doublet remains degenerate. Keeping only a momentum-independent mass instead gives constant eigenvectors. These observations identify the roles of angular winding and the gyroscopic term. They are explanations within the reduced theory, not additional lattice phases or assignments of Chern numbers to touching bands.

\subsection{Distributing the star conversion over both triangle types}
To test the role of triangle asymmetry, replace only the star term by
\begin{equation}
H_J=-\sum_{\tau\in A\cup B}J_\tau\left(\prod_{\alpha\in\tau}d_\alpha+\mathrm{H.c.}\right),\qquad
J_\Sigma=J_A+J_B=0.5.
\label{eq:Sstarcontrol}
\end{equation}
We interpolate from $(J_A,J_B)=(0.5,0)$ to $(0.25,0.25)$ at both mixing strengths and both hopping ratios $t_A/\lambda=0.8,0.5$. Table~\ref{tab:Sstarcontrol} lists three reference allocations; the continuous corner-gap scan and transition are given below. The uniform phase potential and saddle depend on $J_\Sigma$, giving the same chiral vacua and $r=\lambda-7.5$. The maximum complex saddle residual is $1.59\times10^{-15}$. The Bloch normal block is unchanged, while the off-diagonal anomalous block becomes
\begin{equation}
B_{\mu\nu}(\mathbf k)=V\rho^2(1+z_{\mu\nu})-\rho(J_A+J_Bz_{\mu\nu}),\qquad\mu\ne\nu.
\end{equation}
At $\Gamma$, both phase poles remain Eq.~(6) of the Letter.

The coherent-state and positivity bounds extend directly. Write the generalized Hamiltonian as a convex combination of the models with all star conversion on $A$ or on $B$, keeping all other couplings and $J_\Sigma$ fixed. Both have the same coherent minimum under Eq.~\eqref{eq:Scondition} with $J=J_\Sigma$. Their quadratic lower bounds use the corresponding local hopping. Thus
\begin{equation}
\mathcal H_{\rm BdG}(\mathbf k)\succeq
\left[\frac{J_A}{J_\Sigma}\varepsilon(t_A)+\frac{J_B}{J_\Sigma}\varepsilon(t_B)\right]I_6,
\end{equation}
where $\varepsilon(t)$ is the preceding local bound with $m_A$ replaced by $3t/2$. This proves positivity over the full Brillouin zone along the entire interpolation.

An independent real-space bond construction agrees with the Bloch blocks to $3.60\times10^{-15}$. Directional finite differences of the coherent energy converge to the quadratic Hessian, with maximum discrepancy $7.65\times10^{-7}$ at step $2.5\times10^{-3}$. Chern numbers agree on $72^2$, $144^2$, and $288^2$ periodic meshes, in the sewn embedded basis, and with sign reversal under chirality reversal. A nonuniform mesh concentrated near $\Gamma,K,K'$ resolves the narrow curvature peaks: its maximum plaquette phase is below $0.104$ radians and it gives the same integers. Local gap minimizations refine the sampled direct gaps in Table~\ref{tab:Sstarcontrol}; these are numerical minima, distinct from the analytic positivity bound.

\begin{table}[t]
\caption{Stationary star-allocation controls at $\sigma=+1$. The bound $\varepsilon_{AB}$ applies at every momentum; direct gaps are refined numerical minima. $J_A+J_B=0.5$ throughout.}\label{tab:Sstarcontrol}
\begin{ruledtabular}\begin{tabular}{ccccccc}
$\lambda$ & $t_A/\lambda$ & $J_B$ & $\varepsilon_{AB}/\lambda$ & $\Delta_{12}^{\rm dir}/\lambda$ & $\Delta_{23}^{\rm dir}$ & $(C_1,C_2,C_3)$\\
$10^{-4}$ & 0.8 & 0 & 1.199971 & 2.999932 & $0.5207035$ & $(-1,+1,0)$\\
$10^{-4}$ & 0.8 & 0.125 & 0.974978 & 3.000000 & $0.2599006$ & $(-1,+1,0)$\\
$10^{-4}$ & 0.8 & 0.25 & 0.749985 & 3.000000 & $5.474395\times10^{-5}$ & $(-1,0,+1)$\\
$10^{-4}$ & 0.5 & 0 & 0.749989 & 2.999947 & $0.520799$ & $(-1,+1,0)$\\
$10^{-4}$ & 0.5 & 0.125 & 0.749989 & 3.000000 & $0.2599959$ & $(-1,+1,0)$\\
$10^{-4}$ & 0.5 & 0.25 & 0.749989 & 3.000000 & $0.00015$ & $(-1,0,+1)$\\
$10^{-3}$ & 0.8 & 0 & 1.199712 & 2.999320 & $0.5184672$ & $(-1,+1,0)$\\
$10^{-3}$ & 0.8 & 0.125 & 0.974780 & 3.000000 & $0.2576793$ & $(-1,+1,0)$\\
$10^{-3}$ & 0.8 & 0.25 & 0.749847 & 3.000000 & $0.0005474448$ & $(-1,0,+1)$\\
$10^{-3}$ & 0.5 & 0 & 0.749888 & 2.999473 & $0.5194217$ & $(-1,+1,0)$\\
$10^{-3}$ & 0.5 & 0.125 & 0.749888 & 3.000000 & $0.2586323$ & $(-1,+1,0)$\\
$10^{-3}$ & 0.5 & 0.25 & 0.749888 & 3.000000 & $0.0015$ & $(-1,0,+1)$\\
\end{tabular}\end{ruledtabular}\end{table}

\begin{figure}[t]
\centering
\includegraphics[width=0.94\textwidth]{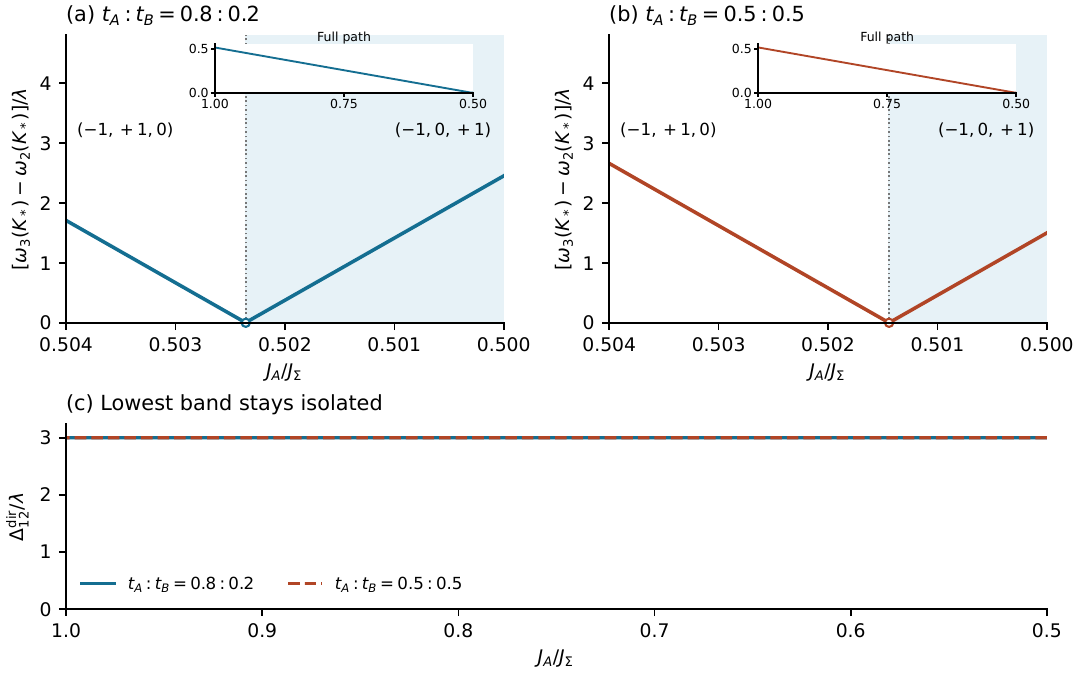}
\caption{Stable upper-band transition along the star interpolation at $\lambda=10^{-3}$ and $\sigma=+1$. (a,b) The corner separation $\omega_3(\mathbf K_*)-\omega_2(\mathbf K_*)$ closes before equal stars, for the two hopping ratios shown. The main axes resolve the transition; insets cover the full allocation path. The horizontal axis decreases toward the symmetric endpoint. Labels give full-zone Chern vectors on either side. (c) Numerically sampled lower direct gap along the same path, normalized by $\lambda$, including additional sampling near $\Gamma$; it remains open. Gaussian stability holds throughout by the analytic positive bound. The corner separation in (a,b) is not asserted to be the global minimum away from the touching.}\label{fig:SstarTransition}
\end{figure}

The lowest-band topology survives the interpolation to equal stars. The change in the full Chern vector from $(-1,+1,0)$ to $(-1,0,+1)$ occurs through a finite-frequency touching of bands 2 and 3,
\begin{equation}
\omega_2(\mathbf K_*)=\omega_3(\mathbf K_*),\qquad
\mathbf K_*=(2\mathbf b_1+\mathbf b_2)/3,
\label{eq:Sstartouch}
\end{equation}
before $J_A=J_B$. Figure~\ref{fig:SstarTransition} resolves the closing and reopening; Table~\ref{tab:Sstarcritical} gives the critical allocations for both mixing strengths. At the touching, $\omega_{2,3}\simeq13.7$, while the remaining numerical separation is below $1.8\times10^{-14}$. The upper-band Chern transfer takes place entirely within the stable positive-frequency sector and is not accompanied by a loss of Gaussian stability. The positive bound above holds at the crossing as well as on either side.

At each zone corner, the normal, anomalous and hole blocks are simultaneously diagonal in the three-component Fourier basis. Keeping these rotation labels fixed allows a signed frequency difference to be followed through zero, rather than minimizing an avoided-crossing fit. A bracketed root is checked against the full six-component Cholesky spectrum and direct diagonalization of $\Sigma_z\mathcal H_{\rm BdG}$. The corner scan covers the full star interpolation and resolves the narrow transition region separately. For the lower direct gap, 65 evenly spaced star allocations plus each critical and bracketing allocation are checked on a $72^2$ zone mesh, supplemented by 70 logarithmic radii from $10^{-7}$ to $0.3$ and 48 angles near $\Gamma$. The smallest sampled $\Delta_{12}^{\rm dir}/\lambda$ exceeds $2.9993$ across both mixing strengths and hopping ratios. These numerical isolation checks complement the denser meshes at the three reference allocations and on either side of the upper-band transition.

For each root $J_B^c$, Chern calculations at $J_B=J_B^c\pm(0.25-J_B^c)/2$ give $(-1,+1,0)$ below and $(-1,0,+1)$ above. Nonuniform $336^2$ and $671^2$ meshes concentrated near $\Gamma$ and both zone corners agree; their maximum plaquette phases are below $0.078$ and $0.025$ radians, respectively. The touching bands' individual Chern numbers are undefined at the root; their combined value remains $C_2+C_3=+1$. The lowest band remains isolated with $C_1=-1$. Thus the opposite-Chern low doublet belongs to the original benchmark completion, while the topological lowest pseudo-Goldstone band persists at the fully symmetric endpoint.

\begin{table}[t]
\caption{Upper-band transition at $\sigma=+1$, $J_A+J_B=0.5$. Frequencies are in $t'=1$. The tabulated digits locate the crossing of the specified Gaussian Hamiltonian.}\label{tab:Sstarcritical}
\begin{ruledtabular}\begin{tabular}{ccccc}
$\lambda$ & $t_A/\lambda$ & $J_A^c$ & $J_B^c$ & $\omega_2=\omega_3$\\
$10^{-4}$ & 0.8 & 0.250117895 & 0.249882105 & 13.7000331\\
$10^{-4}$ & 0.5 & 0.250072105 & 0.249927895 & 13.6999881\\
$10^{-3}$ & 0.8 & 0.251179008 & 0.248820992 & 13.7011521\\
$10^{-3}$ & 0.5 & 0.250721090 & 0.249278910 & 13.7007024\\
\end{tabular}\end{ruledtabular}\end{table}

The conserving-parent polarization retains winding two. Distributing the star term also changes its gradient dynamics: eliminating the common phase gives sound speeds
\begin{equation}
c_T^2=3t'a_v,\qquad c_L^2=3t'a_v+a_v\rho\frac{J_AJ_B}{J_\Sigma},\qquad
a_v=J_\Sigma\rho+2(u-V)\rho^2.
\end{equation}
For the three star allocations, $(c_T,c_L)$ approach $(2.738613,2.738613)$, $(2.738613,2.781074)$, and $(2.738613,2.795085)$. Reducing $q$ from $0.04$ to $0.005$ brings the computed phase velocities within $1.24\times10^{-5}$ of these limits. The phase Stokes vector $(|p_x|^2-|p_y|^2,2\mathrm{Re}(p_x^*p_y))/|\mathbf p|^2$ winds twice on the tested parent circles $q=0.04,0.08,0.16$. With nonzero $J_AJ_B$, unequal sound speeds produce a leading linear frequency splitting; the cubic splitting and $q_*\propto\lambda^{1/3}$ of the Letter refer to its $J_B=0$ benchmark. The finite-resolution analysis also uses that original benchmark.

\section{Canonical phase dynamics, low-energy matching, and topology}
\subsection{Gyroscopic term and phase pinning}
Using $Q=(a+a^\dagger)/\sqrt2$, $P=(a-a^\dagger)/(i\sqrt2)$ and projecting onto $E$, the relative sector at $\Gamma$ is
\begin{equation}
H_v=\frac12Q^TXQ+Q^TCP+\frac12P^TYP,\quad
X=(2a_v+m)I_2,\quad Y=mI_2,\quad C=\sigma mR,
\quad R=\begin{pmatrix}0&-1\\1&0\end{pmatrix}.
\end{equation}
Starting from $L=P^T\dot Q-H_v$ and eliminating $Q$ gives
\begin{equation}
L_v=\frac12(\dot P+CP)^TX^{-1}(\dot P+CP)-\frac12P^TYP.
\end{equation}
With $P=\sqrt2\rho\varphi$, this yields the phase action in the Letter, $\chi_\lambda=2\rho^2/(2a_v+m)$ and $\mu^2=2a_vm$. The two poles are $\sqrt{\mu^2+m^2}\pm m$; the total phase mode has $\omega_s(0)=\sqrt{3jX_s}$, where $X_s=4(u+2V)\rho^2-j$. The antisymmetric amplitude--phase block is essential for the circular splitting. Counting phase-Hessian zero directions alone does not give these frequencies~\cite{WM2012}.

\subsection{Conserving-parent polarization of the benchmark}
For the original $J_B=0$ completion at $t_\Sigma=0$, let $X=A+B$ and $Y=A-B$. In the periodic basis,
\begin{align}
Y&=\gamma I_3+j\mathbf s\mathbf s^\dagger,\\
X&=(\gamma+2a_v)I_3+\alpha\mathbf s\mathbf s^\dagger+b\mathbf v\mathbf v^\dagger,\quad
\alpha=2V\rho^2-j,\quad b=2V\rho^2,\quad v_\nu=e^{2i\mathbf q\cdot\delta_\nu}.
\end{align}
Squared frequencies are eigenvalues of $Y^{1/2}XY^{1/2}$. This is a scalar matrix plus a rank-two update; one branch is exactly $\omega_T^2=\gamma(\gamma+2a_v)$. Since $\gamma=gq^2+O(q^4)$ with $g=3t'/2$, eliminating the gapped total-phase sector gives, in the low-frequency eigenproblem,
\begin{equation}
X_{v,\rm eff}=(2a_v+\gamma)I_2+\beta\mathbf q\mathbf q^T+\cdots,
\quad \beta=V\rho^2\left(1-\frac{6V\rho^2}{X_s}\right).
\end{equation}
Both sound speeds are $c=\sqrt{3t'a_v}$, but $\omega_L-\omega_T=2aq^3+O(q^5)$, with $a=3t'\beta/(8c)$. Their polarization projectors contain $\cos2\phi$ and $\sin2\phi$. For the benchmark, $a_v=2.5$, $X_s=15.5$, $\beta=19/31$, and $a=0.0839252306$.

In the joint Gaussian limit $q\sim\lambda^{1/3}\gg\lambda^{1/2}$, the positive-frequency projected matrix is
\begin{equation}
h_{\rm low}-\bar\omega I_2=aq^3(\cos2\phi\,\sigma_z+\sin2\phi\,\sigma_x)-\sigma m\sigma_y+o(\lambda).
\end{equation}
The factor $q=|\mathbf q|$ is a consequence of projection onto positive acoustic frequencies, not a nonlocal microscopic term. In a general symmetry-allowed effective action $\kappa_L$ and $\kappa_T$ need not be equal. Thus neither the $q^3$ splitting nor its radial scaling follows from $U(1)^2$ alone.

\subsection{Phase polarization in the full lattice model}
The circularity plotted in Fig.~2(b) of the Letter is evaluated from the phase component of the normalized positive-frequency BdG eigenvector in the periodic cell basis. With $w_1=(u_1,v_1)^T$, define
\begin{equation}
\mathbf p_1=E^T(u_1-v_1),\qquad
P_{\rm circ}=\frac{2\operatorname{Im}(p_{1x}^*p_{1y})}{|\mathbf p_1|^2}.
\end{equation}
The common phase and normalization of $w_1$ cancel. Circular and linear polarizations have $|P_{\rm circ}|=1$ and $P_{\rm circ}=0$, respectively. The two-mode Hamiltonian gives $P_{\rm circ}\simeq\sigma m/\sqrt{m^2+f(q)^2}$; the Letter compares this with the full lattice result along $q\hat x$ at $\lambda=10^{-4}$, $\sigma=+1$. Finite-momentum corrections become visible beyond the crossover. This phase-space circularity is defined in the cell convention of the tensor source; the Berry curvature is calculated independently in the embedded orbital convention. 

\subsection{Local winding contribution versus an absolute Chern number}
For the explicitly defined two-band matrix
\begin{equation}
h-\bar\omega I_2=f(q)[\cos(w\phi)\sigma_z+\sin(w\phi)\sigma_x]-M\sigma_y,
\end{equation}
with signed $M\ne0$, integer winding $w$, and $f(0)=0$, the lower-band curvature and disk flux are
\begin{align}
\Omega_-(q)&=-\frac{wMff'}{2q(M^2+f^2)^{3/2}},\\
\frac{\Phi_-(R)}{2\pi}&=-\frac w2\left[\operatorname{sgn}M-\frac{M}{\sqrt{M^2+f(R)^2}}\right].
\label{eq:Sflux}
\end{align}
This follows by differentiating $M/\sqrt{M^2+f^2}$. If $|f(R)|\gg|M|$, the singular local contribution tends to $-w\operatorname{sgn}M/2$. For $w=2$ a single side contributes one unit in magnitude; under a common ultraviolet completion with no other closings, the difference $C_-(-|M|)-C_-(+|M|)$ is instead $w=2$.

The local expression does not fix the absolute global invariant. For example $f=q^w$, $M(q)=m-Bq^{2w}$ has a compactified-plane Chern number
\begin{equation}
C_-=-\frac w2[\operatorname{sgn}m+\operatorname{sgn}B].
\end{equation}
At $w=2,m>0$, the choices $B>0$ and $B<0$ give $-2$ and $0$, despite the same small-disk contribution $-1$. The winding--mass mechanism~\cite{Sun2009} acts here within the symmetry-protected dipole phase sector. The star interpolation makes the distinction between local geometry and global Chern allocation concrete: the lowest-band Chern number survives, while the compensating unit moves between higher bands through Eq.~\eqref{eq:Sstartouch}.

The parent modes touch their negative-frequency partners at the origin. For the benchmark, applying the local result to this bosonic problem requires a matching annulus $q_{\rm pg}\sim\lambda^{1/2}\ll q_*\sim\lambda^{1/3}\ll R\ll1$ and an independent lattice calculation. The full-BdG calculation supplies the latter. For the Gaussian two-mode form,
\begin{equation}
\Omega_1(q)\simeq-\sigma\frac{3ma^2q^4}{(m^2+a^2q^6)^{3/2}},\quad
q_*=(m/|a|)^{1/3},\quad
-\sigma q_*^2\Omega_1=\frac{3(q/q_*)^4}{[1+(q/q_*)^6]^{3/2}}.
\end{equation}
The peak radius scales as $\lambda^{1/3}$ and the peak height as $\lambda^{-2/3}$, within this specified Gaussian limit.

\section{Gaussian occupation and the control limit}
The fluctuation occupation per three-orbital cell is
\begin{equation}
n_{\rm ex}=\frac1{|\mathrm{BZ}|}\int d^2k\sum_{n>0,\nu}|v_{\nu n}(\mathbf k)|^2.
\end{equation}
Because total dipole number is not conserved, $n_{\rm ex}$ means occupation outside the coherent background, not a fixed-number subtraction. In the parent the low-energy integrand behaves as $a_v/(cq)$ and the two-dimensional zero-temperature integral is infrared finite. With weak pinning,
\begin{equation}
n_{\rm ex}(0)-n_{\rm ex}(t_\Sigma)
=\frac{\sqrt{a_v}}{4\pi t'}\sqrt{t_\Sigma}+o(\sqrt{t_\Sigma}).
\end{equation}
\begin{table}[t]
\caption{Quantum-occupation results on the constant-amplitude slice.}
\begin{ruledtabular}
\begin{tabular}{ccc}
$\lambda$&$n_{\rm ex}$ per cell&$n_{\rm ex}/(3\rho^2)$\\
$10^{-4}$&0.2227059201&0.0742353067\\
$10^{-3}$&0.2200256078&0.0733418693
\end{tabular}
\end{ruledtabular}
\end{table}
We integrate all six hexagonal-BZ sectors with polar Gauss--Legendre quadrature. The three grids have $(N_r,N_\phi)=(64,24)$, $(128,48)$, and $(256,64)$ per sector. The last changes in $n_{\rm ex}$ are $1.40\times10^{-11}$ and $3.1\times10^{-16}$ at the two benchmarks. A scalar Cholesky calculation verifies the batch integrand at selected momenta. These checks quantify numerical integration accuracy within the Gaussian approximation.

The equality of the leading longitudinal and transverse stiffnesses in the Gaussian benchmark is not enforced by dipole symmetry. Interaction corrections can generate a lower-order splitting, so the cubic splitting and its radial scaling apply within the stated Gaussian matching regime.

A systematic family exists without changing the form of Eq.~\eqref{eq:Smodel}: take $\rho\mapsto\sqrt{\mathcal N}\rho$, $(u,V)\mapsto(u,V)/\mathcal N$, $J\mapsto J/\sqrt{\mathcal N}$, and keep $r,t',t_A,t_B$ fixed. The quadratic spectrum is unchanged, the coherent energy is $O(\mathcal N)$, cubic and quartic fluctuation vertices are $O(\mathcal N^{-1/2})$ and $O(\mathcal N^{-1})$, and relative occupation/lowest-order rates scale as $1/\mathcal N$. The reported baseline is $\mathcal N=1$. Small occupation alone does not bound relative errors in an arbitrarily small band splitting.

\section{Tensor source, torque identity, and finite-momentum spectroscopy}
\subsection{Specified source protocol and exact linear current}
The external field is introduced only on same-flavor hopping, as
\begin{equation}
H_{t'}[a]=-t'\sum_{\mathbf R,\nu,\boldsymbol\ell}
\left[e^{iE_{\nu a}\int_0^1ds\,\ell_i a_{ia}(\mathbf R+s\boldsymbol\ell,t)}
 d_{\mathbf R+\boldsymbol\ell,\nu}^\dagger d_{\mathbf R,\nu}+\mathrm{H.c.}\right].
\end{equation}
The source is defined at common cell coordinates, not silently substituted by a field at literal orbital centers. This makes the protocol reproducible while avoiding an unproved microscopic electromagnetic mapping. Rank-2 synthetic-field proposals provide related context, not a ready-made implementation of the present lattice~\cite{Zhang2026}.

Differentiating gives
\begin{align}
J_i^{a(1)}(\mathbf q)&=s_i(\mathbf q)\sum_\nu E_{\nu a}(a_{\mathbf q\nu}-a_{-\mathbf q\nu}^\dagger),\\
s_i(\mathbf q)&=2t'\rho\sum_{\boldsymbol\ell}\ell_i
\sin(\mathbf q\cdot\boldsymbol\ell/2)
\frac{\sin(\mathbf q\cdot\boldsymbol\ell/2)}{\mathbf q\cdot\boldsymbol\ell/2}.
\end{align}
The ratio equals one at zero argument. Its sinc factor is fixed by the line-integrated source. The long-wave limit is $J_i^{a(1)}\simeq i\kappa q_i\varphi_a$. Its phase vertex is distinct from the density vertex $F_a=\rho\sum_\nu E_{\nu a}(a_\nu+a_\nu^\dagger)$. At exactly $q=0$ the one-quasiparticle current vertex vanishes.

In the $E$ normalization, physical $\mathbf e$-weighted currents are $J_{\rm phys}=J/\sqrt2$ and their weights and response kernels are half the values tabulated here. The distinction is a normalization, not an additional effect.

An explicit linear lattice check uses $l_F=\rho(E^T,E^T)$, $\mathcal D=\Sigma_z\mathcal H$, and $l_{J_i}=s_i(E^T,-E^T)$. Let $h_{\rm mix}$ have zero diagonal and off-diagonal entries $(t_A+t_Bz_{\mu\nu})e^{i(\theta_\nu-\theta_\mu)}$. Then
\begin{equation}
l_F\mathcal D-q_i l_{J_i}
=\rho\bigl(E^T[h_{\rm mix}(\mathbf q)+t_\Sigma I],
-E^T[h_{\rm mix}^{T}(-\mathbf q)+t_\Sigma I]\bigr).
\label{eq:SlatticeWard}
\end{equation}
The right side is the linear row for $[F,H_{\rm mix}]=i\tau$. The $t_\Sigma I$ terms include the condensate-background contribution; commuting only linear $F$ with quadratic hopping would omit them.

\subsection{Sources, explicit nonconservation, and the response kernel}
Introduce temporal/spatial sources $a_0,a_i$ and a fixed breaking reference phase $\zeta$. The local quadratic action is
\begin{equation}
\mathcal L[a,\zeta]=\frac{\chi_\lambda}{2}
|\dot\varphi-a_0+C(\varphi-\zeta)|^2
-\frac{\chi_\lambda(2a_v+m)m}{2}|\varphi-\zeta|^2
-\frac\kappa2|\partial_i\varphi-a_i|^2+\cdots.
\end{equation}
It is covariant under $\varphi,\zeta\mapsto\varphi+\beta,\zeta+\beta$ and $a_\mu\mapsto a_\mu+\partial_\mu\beta$. The spurion $\zeta$ is not a new dynamical field. Defining $F=\delta\mathcal L/\delta a_0=-\pi_\varphi$, $J_i=\delta\mathcal L/\delta a_i$, and $\tau=\delta\mathcal L/\delta\zeta$ gives
\begin{equation}
\partial_tF_a+\partial_iJ_i^a=\tau_a.
\end{equation}
The physical system at finite mixing has nonzero torque; source covariance does not restore its exact conservation.

A scalar-charge rank-2 source is a restricted embedding: $a_{0a}=\partial_a\mathsf A_0/\sqrt2$, $a_{ia}=\mathsf A_{ia}/\sqrt2$ with $\mathsf A_{ia}=\mathsf A_{ai}$, and $\beta_a=\partial_a\alpha/\sqrt2$. It is a restriction of two independent background $U(1)$ sources, not identical to their most general coupling.

For $e^{i\mathbf q\cdot\mathbf r-izt}$ and $z=\omega+i0^+$,
\begin{equation}
G^{-1}=\chi_\lambda[(\mu^2+c_\lambda^2q^2-z^2)I_2-2i\sigma mzR],\qquad
\Pi_{ia,jb}=-\kappa\delta_{ij}\delta_{ab}+\kappa^2q_iq_jG_{ab}.
\end{equation}
The contact term is essential. With $A=\mu^2+c_\lambda^2q^2-z^2$, the chirality-odd part is
\begin{equation}
\Pi^{\rm odd}_{ia,jb}=\frac{2i\sigma\kappa^2mzq_iq_j}{\chi_\lambda(A^2-4m^2z^2)}R_{ab}.
\end{equation}
It is not a quantized dc Hall coefficient. For the retarded convention $\chi^R=-i\Theta(t)\langle[J(t),J(0)]\rangle$ and perturbation $H'=-aJ$, the source response is $\Pi=-K_{\rm dia}-\chi^R$. This fixes an otherwise ambiguous overall sign. The exact lattice torque identity, source derivative, and Onsager--Casimir checks are summarized below.

\subsection{Circular tensor weights and geometry reconstruction}
Let
\begin{equation}
M_\pm=\frac12\begin{pmatrix}1&\pm i\\\pm i&-1\end{pmatrix},\quad
O_\pm=M_{\pm,ia}J_i^a,
\quad W_{\pm,n}=|\langle0|O_\pm|n,\mathbf q\rangle|^2.
\end{equation}
For a drive $H'=-\mathcal A e^{-i\omega t}O_\pm^\dagger+\mathrm{H.c.}$, these weights define the positive-frequency absorption. Reversing the complex-drive convention exchanges labels. At $q\parallel x$, $q\to0$, and $\sigma=+1$, $O_-$ selects the lower branch and $O_+$ the second branch, with allowed weight
\begin{equation}
W_{\rm allowed}=\frac{\kappa^2q^2}{4\chi_\lambda\sqrt{m(2a_v+m)}}+o(q^2).
\end{equation}
The reversed chirality exchanges the rule. At $q=10^{-5}$, the computed $W/q^2$ is 205.3989373 for $\lambda=10^{-4}$ and 64.9616441 for $10^{-3}$; the corresponding limiting predictions are 205.3990400 and 64.9616473. This is a finite-$q$ current measurement, not the uniform density measurement discussed below.

Normalize the contrast at each angle before averaging:
\begin{equation}
\mathcal D_1(q,\phi)=\frac{W_{+,1}-W_{-,1}}{W_{+,1}+W_{-,1}},\qquad
\overline{\mathcal D}_1(q)=\frac1{2\pi}\int_0^{2\pi}d\phi\,\mathcal D_1(q,\phi).
\end{equation}
Define $\overline{\Omega}_1(q)=(2\pi)^{-1}\int_0^{2\pi}d\phi\,\Omega_1(q,\phi)$ in the embedded orbital convention. In the isotropic winding-two low-energy limit,
\begin{equation}
\overline{\mathcal D}_1\simeq-\frac{\sigma m}{\sqrt{m^2+f(q)^2}},\quad
\overline{\Omega}_1\simeq-\frac1q\frac{d\overline{\mathcal D}_1}{dq},\quad
\frac{\Phi_1(R)}{2\pi}\simeq-\sigma-\overline{\mathcal D}_1(R).
\end{equation}
This is a model-assisted local relation, not a universal full-BZ sum rule. Quantum geometry and dichroic probes have established precedents~\cite{Tran2017,Tesfaye2025}; here the source vertex and doublet structure make a specific connection that can be tested separately against the full BdG curvature.

\begin{figure}[htbp]
\centering
\includegraphics[width=0.65\textwidth]{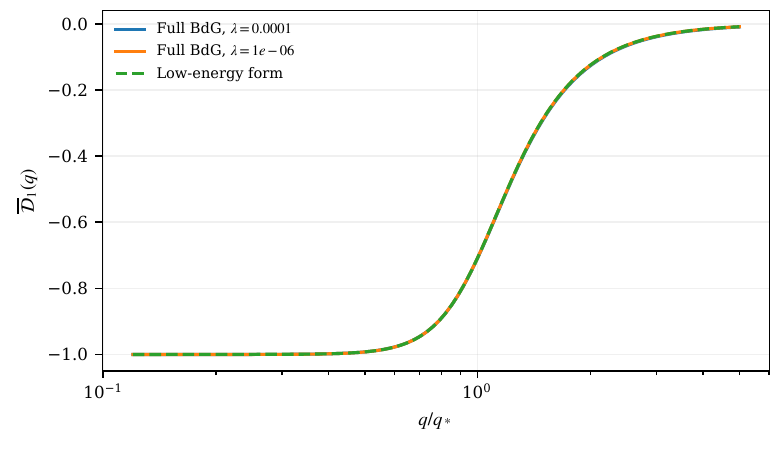}
\caption{Ideal angular-averaged tensor circular dichroism from the six-dimensional BdG data. The dashed curve is $-1/\sqrt{1+(q/q_*)^6}$. The comparison concerns a low-energy source-defined contrast, not an absolute full-band Chern measurement.}
\label{fig:Scontrast}
\end{figure}

\begin{table}[t]
\caption{Independent comparisons at $q=q_*$. The final two columns use distinct calculations: a full symplectic curvature formula and derivatives of the spectral contrast. The small-$\lambda$ rows test Gaussian matching, not the interacting infrared limit.}
\begin{ruledtabular}
\begin{tabular}{cccc}
$\lambda$&$\overline{\mathcal D}_1(q_*)$&$-q_*^2\overline{\Omega}_1$&Spectral reconstruction\\
$10^{-4}$&$-0.7082830$&1.0550449&1.0546989\\
$10^{-5}$&$-0.7072254$&1.0593743&1.0592957\\
$10^{-6}$&$-0.7070691$&1.0603512&1.0603323\\
$10^{-7}$&$-0.7070693$&1.0605789&1.0605832
\end{tabular}
\end{ruledtabular}
\end{table}
The comparisons use 24 angles and central radial differences of step $10^{-3}q_*$. The limiting values are $-1/\sqrt2$ and $3/(2\sqrt2)$. At $R=4q_*$, spectral flux estimates approach the two-mode result $-0.9843769$. Figure~\ref{fig:Scontrast} shows the ideal contrast. High-mode admixture, lattice anisotropy, orbital connections, and probe calibration produce corrections. These ideal spectra precede instrumental broadening and noise; their wider momentum range extends beyond the tested decay-closed window in the Letter.

\subsection{Common benchmark window and reproducible checks}
Figure~4 of the Letter uses both $\lambda=10^{-4}$ and $10^{-3}$ over $0.35\le q/q_*\le1.1$. Table~\ref{tab:Scommon} collects the same-parameter evidence. We independently compute the source weights and the full six-mode embedded curvature, without using the two-mode prediction. The radial grid contains 31 points, with 24 angles at each radius and derivative step $10^{-3}q_*$. At $q_*$, doubling the angular grid and halving the radial step changes the reconstructed scaled curvature by less than $1.9\times10^{-6}$; this numerical change is distinct from the mismatch of the low-energy reconstruction to the full curvature.

For threshold checks we used $q/q_*=0.35,0.5,0.75,1,1.1$ and twelve angles $\phi=2\pi j/12$. At each momentum, a $12\times12$ reciprocal-parallelogram search supplies the five best starting points; equal partition and both endpoints supply three additional starts. Each is refined by Nelder--Mead in the full two-dimensional momentum plane, using periodic lattice frequencies. All optimizers converged with coordinate tolerance $10^{-9}$ and energy tolerance $10^{-12}$. Positive margins are numerical evidence at these samples, not a rigorous lower bound over all final or initial momenta. The onset checks along $\hat x$ put the first opening of either low-band channel at $q/q_*=1.4272$ and $1.1380$, respectively, beyond the displayed endpoint.

\begin{table}[t]
\caption{Same-parameter Gaussian benchmarks. Weights and energies are evaluated at $\mathbf q=q_*\hat x$, $\sigma=+1$, in the $E$ source normalization. Curvature entries are angular averages. Direct gaps are sampled $144^2$-grid minima. Threshold minima include both initial low bands and the stated angular samples. Reconstruction mismatch is relative to the independently computed embedded curvature.}\label{tab:Scommon}
\begin{ruledtabular}
\begin{tabular}{lcc}
Quantity & $\lambda=10^{-4}$ & $\lambda=10^{-3}$\\
$q_*$ & 0.12135739 & 0.26145658\\
$\Delta_{12}^{\rm dir}$ & 0.00030000 & 0.00299961\\
$\Delta_{23}^{\rm dir}$ & 0.52070352 & 0.51846720\\
$(\omega_1,\omega_2)$ & (0.33400033, 0.33442814) & (0.72647326, 0.73076209)\\
$(W_{+,1},W_{-,1})$ & (0.03412969, 0.21453044) & (0.06692254, 0.46855085)\\
$(W_{+,2},W_{-,2})$ & (0.21461350, 0.03420498) & (0.46939101, 0.06760276)\\
$\overline{\mathcal D}_1(q_*)$ & $-0.70828301$ & $-0.71373478$\\
$-q_*^2\overline{\Omega}_1(q_*)$ & 1.05504491 & 1.03606070\\
Spectral reconstruction at $q_*$ & 1.05469881 & 1.03465701\\
Relative mismatch at $q_*$ & $0.0328\%$ & $0.1355\%$\\
Largest relative mismatch on 31-point grid & $0.401\%$ & $1.861\%$\\
Smallest threshold margin at $q_*$ & 0.00229404 & 0.00488638\\
Smallest threshold margin on five-radius grid & 0.00177356 & 0.00138804
\end{tabular}
\end{ruledtabular}
\end{table}

The spectra in Fig.~4(a,b) use the Gaussian convolution defined below, a common reference splitting, and one shared weight normalization for the two helicities. At fixed momentum the opposite chiralities need not have identical poles, since time reversal relates $(\sigma,\mathbf q)$ to $(-\sigma,-\mathbf q)$. Each independently calculated spectrum is centered on its own mean frequency. Both panels use the same instrumental widths, reference splitting, and intensity normalization.

Independent numerical checks test the lattice torque identity, Eq.~\eqref{eq:SlatticeWard}, the $\Gamma$ poles, and the large-occupation matrix scaling. The Ward residual is $1.17\times10^{-15}$ and the $\Gamma$ frequency error is $2.42\times10^{-14}$.

\subsection{Finite frequency and momentum resolution}
The ideal vertex determines the zero-temperature one-quasiparticle spectrum
\begin{equation}
 S_h(\omega,\mathbf q)=\sum_n W_{h,n}(\mathbf q)\delta[\omega-\omega_n(\mathbf q)],\qquad h=\pm.
\end{equation}
We model instrumental resolution by a normalized Gaussian of standard deviation $s_\omega$ in frequency and an isotropic two-dimensional Gaussian with component standard deviation $s_q$ in momentum. Thus
\begin{equation}
 \widetilde S_h(\omega,\mathbf q)=\int d^2k\,G_{s_q}(\mathbf k-\mathbf q)\sum_n W_{h,n}(\mathbf k)\,g_{s_\omega}[\omega-\omega_n(\mathbf k)].
\end{equation}
The zero-width limits are delta distributions. These Gaussians describe a synthetic instrument, not an interaction self-energy. The intrinsic poles remain sharp by the Gaussian approximation, not by an all-orders lifetime proof. No background counts, calibration drift, or thermal broadening are included.

At $\mathbf q=q_*\hat x$ and $\sigma=+1$, define $\Delta\omega_*=\omega_2-\omega_1$, $\bar v_*=|\partial_q(\omega_1+\omega_2)/2|$, and
\begin{equation}
 q_{\rm res}=\Delta\omega_*/\bar v_*.
\label{eq:SresolutionScale}
\end{equation}
This momentum scale compares the frequency shift across a momentum bin with the doublet separation; it is not a universal inference threshold. We use all 25 combinations of $s_\omega/\Delta\omega_*=0,0.1,0.25,0.5,1$ and $s_q/q_{\rm res}=0,0.1,0.3,1,3$. Numerical scales are listed in Table~\ref{tab:SresolutionScales}. For the displayed setting $s_q=0.1q_{\rm res}$, the required ratios are $s_q/q_*=1.28\times10^{-4}$ and $5.86\times10^{-4}$ at $\lambda=10^{-4}$ and $10^{-3}$, respectively. Since $\Delta\omega_*\sim\lambda$ and $\bar v_*$ remains finite, $q_{\rm res}\sim\lambda$ and $q_{\rm res}/q_*\sim\lambda^{2/3}$ within this Gaussian regime. The resolution requirement therefore tightens towards the conserving limit.

\begin{table}[t]
\caption{Resolution reference scales. $q_{\rm res}$ and $q_*$ are in inverse triangular-spacing units, and frequencies are in $t^{\prime}$. The selected main-figure setting uses $s_\omega=0.25\Delta\omega_*$ and $s_q=0.1q_{\rm res}$. All grid widths follow by multiplying these reference scales by the stated ratio lists.}\label{tab:SresolutionScales}
\begin{ruledtabular}\begin{tabular}{cccccc}
$\lambda$ & $q_*$ & $\Delta\omega_*$ & $\bar v_*$ & $q_{\rm res}$ & $q_{\rm res}/q_*$ \\
$1.00000\times10^{-4}$ & $1.21357\times10^{-1}$ & $4.27808\times10^{-4}$ & $2.74740\times10^{0}$ & $1.55714\times10^{-4}$ & $1.28310\times10^{-3}$\\
$1.00000\times10^{-3}$ & $2.61457\times10^{-1}$ & $4.28883\times10^{-3}$ & $2.80114\times10^{0}$ & $1.53110\times10^{-3}$ & $5.85606\times10^{-3}$\\
\end{tabular}\end{ruledtabular}\end{table}

\subsection{Spectral extraction and independent evaluation}
The forward calculation diagonalizes the full six-component BdG matrix at Gaussian-distributed momenta and applies the exact source vertex. Acquisition assumes that the low doublet has been located and the frequency axis calibrated: the window is centered on its mean energy and spans the doublet range plus instrumental broadening. Within this window, the fit receives only the two polarization-resolved frequency-bin arrays, bin boundaries, and one common exposure. Two Gaussian peaks share fitted centers and one width, with four independent nonnegative areas. True mode energies, weights, dichroism, and curvature are reserved for evaluation. Distinct group velocities and non-Gaussian momentum-averaged line shapes can violate the common-width assumption and contribute to the reported bias; high-mode leakage is evaluated separately.

We minimize Poisson deviance over centers, separation, width, and areas using multiple deterministic starts. We retain convergence and boundary flags, residuals, and the fitted separation-to-width ratio; optimizer convergence alone does not establish spectral identifiability. At zero instrumental widths, integrating the two occupied bins recovers their areas directly from the generated spectra.

For each angle, the fitted lower-mode areas give
\begin{equation}
 \widehat{\mathcal D}_1(q,\phi)=\frac{\widehat W_{+,1}-\widehat W_{-,1}}{\widehat W_{+,1}+\widehat W_{-,1}}.
\end{equation}
The two polarizations are never normalized independently. We average this contrast over angle before differentiating radially. The main resolution map concerns local curvature at $q_*$: a cubic polynomial fitted to nine radii in $q/q_*\in[0.94,1.06]$ supplies $d\overline{\widehat{\mathcal D}}_1/d(q/q_*)$. The full-BdG embedded curvature is calculated independently and is used only in the final error calculation. Ten percent relative curvature error is a prespecified display criterion, not an experimental law or a fit acceptance constraint. Curves and errors are retained on both sides of it.

For radial illustrations, we also sample 21 radii from $0.5q_*$ to $1.1q_*$ at the predetermined settings $(s_\omega/\Delta\omega_*,s_q/q_{\rm res})=(0.25,0.1)$ and $(1,3)$. A seven-point degree-three Savitzky--Golay derivative gives the reconstructed curvature. The main figure uses interior derivative points; the complete results include endpoint estimates and the unfavorable setting. These radial illustrations are distinct from the nine-radius local derivative used in the resolution map.

\subsection{Numerical convergence and shot noise}
Momentum convolution uses 4096 antithetic scrambled Sobol normal points, with 12 equally spaced angles and 481 frequency bins. All 25 resolution settings are recomputed with 8192 points, 24 angles, 963 bins, and a halved radial fitting window. The frequency kernel is integrated before internal bins are combined. Further refinement tests sensitive cells as described below; only classifications stable under refinement are interpreted.

For each $\lambda$, one fixed exposure gives $10^5$ expected counts summed over both polarizations in the reference $q_*\hat x$ doublet. The same exposure is used at every angle and radius. This is a per-spectrum reference count, not a $10^5$-count budget for the entire reconstruction. We draw independent Poisson bin counts for three settings, $(0.25,0.1)$, $(0.5,1)$ and $(1,3)$. Each repetition consists of five radii in $[0.94,1.06]q_*$ and six angles, hence 30 two-polarization spectra. There are 100 complete repetitions per setting and $\lambda$. A noiseless calculation with this same reduced sampling separates its discretization bias from shot noise.

Distributions include all finite estimates, including those from unsuccessful fits; failures are counted separately. The central 95\% ranges are Monte Carlo percentiles over repeated synthetic measurements and quantify finite-count fluctuations. Model and calibration uncertainty are separate.

\paragraph{Results and interpretation.}

At zero instrumental widths, the batch implementation recovers the scalar BdG frequencies and weights with maximum absolute differences $5.55\times10^{-17}$ and $2.78\times10^{-16}$. Extracted $\overline{\mathcal D}_1(q_*)$ differs from the 48-angle reference by $4.25\times10^{-9}$ and $9.01\times10^{-8}$. The nine-radius cubic derivative differs from the finer central-difference estimate by $2.51\times10^{-5}$ and $2.33\times10^{-5}$ in scaled curvature. This numerical derivative error is separate from the intrinsic mismatch of the local reconstruction relation with the full BdG geometry.

For $(s_\omega/\Delta\omega_*,s_q/q_{\rm res})=(0.25,0.1)$, the coarse-grid curvature errors are $0.0298\%$ and $0.1136\%$. The refined values are $0.0285\%$ and $0.1127\%$; coarse-to-fine changes are only $0.00126\%$ and $0.000856\%$ of the full BdG value. These percentages describe noiseless expected spectra. They do not include shot noise or unknown intrinsic line shapes.

\begin{figure}[t]\centering

\includegraphics[width=0.94\textwidth]{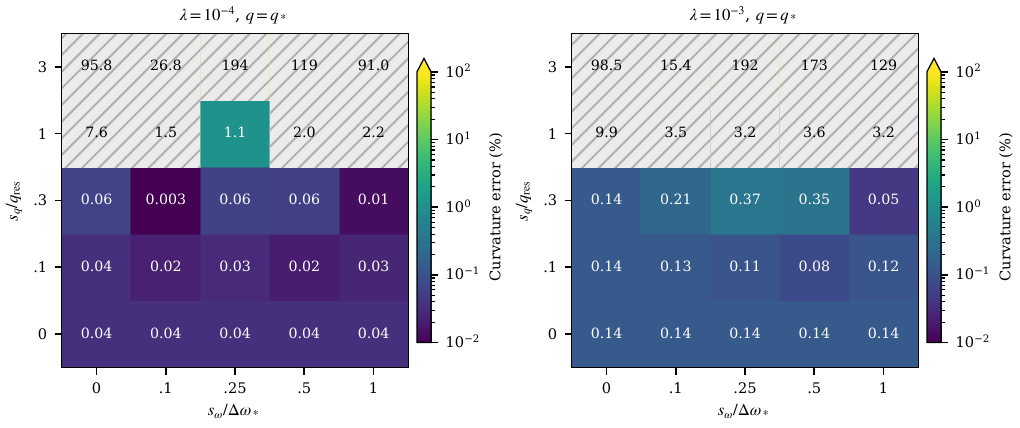}

\caption{Local $q_*$ curvature error in percent after joint spectral extraction. Numbers retain every coarse-grid result. Gray hatching marks numerical sensitivity under the stated refinement and excludes those cells from a reliable classification. Unhatched cells are stable at the stated tolerance; all lie below the prespecified $10\%$ display level. Each row and column changes a fixed absolute instrumental width.}\label{fig:SresolutionMap}\end{figure}

\begin{figure}[t]\centering

\includegraphics[width=0.94\textwidth]{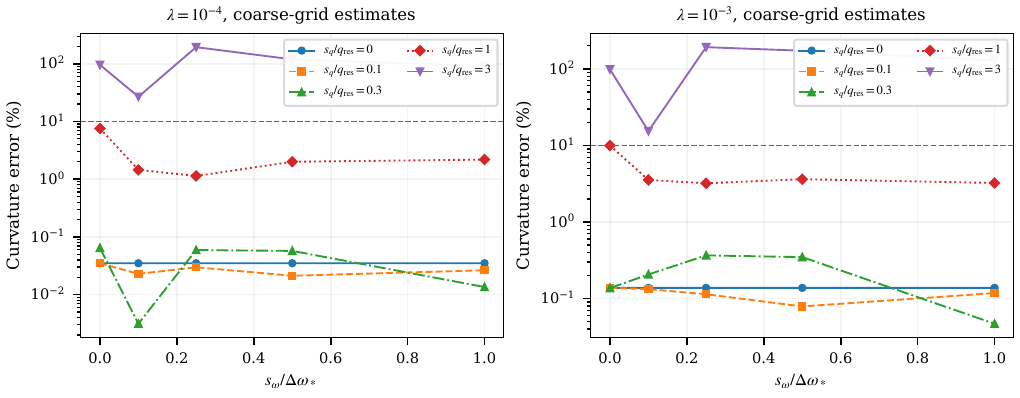}

\caption{The same complete error data as curves against frequency resolution. Symbols and line styles distinguish momentum widths. The horizontal line is the $10\%$ display criterion. Nonmonotonic values in the numerically sensitive region describe the inference algorithm and quadrature, and are not a physical resolution boundary.}\label{fig:SresolutionCurves}\end{figure}

Figures~\ref{fig:SresolutionMap} and \ref{fig:SresolutionCurves} retain every grid setting. Hatching identifies an aggregate-refinement change exceeding $1\%$ of the true curvature or a change across the $10\%$ display threshold. Such cells are numerical-sensitivity flags, not accepted resolution boundaries. The favorable point above is stable under this check. At broad momentum resolution, nearly overlapping lines can give successful optimizer exits while inferred areas and their radial derivatives remain unstable. The stored separation-to-width ratios, Jacobian condition numbers, residuals, and fit-failure counts distinguish these diagnostics.

\begin{table}[t]

\caption{Poisson reconstruction results from 100 complete repeated measurements per setting. Error is signed, $100(\widehat F/F_{\rm BdG}-1)$, where $F=-q_*^2\overline{\Omega}_1$. The central 95\% range includes all finite estimates, including unsuccessful fits. $N_{10}$ counts repetitions within $10\%$ absolute relative error; $N_{\rm fail}$ counts unsuccessful spectral fits out of 3000, not failures of the $10\%$ target.}\label{tab:Snoise}

\begin{ruledtabular}\begin{tabular}{cccccc}

$\lambda$ & $(s_\omega/\Delta\omega_*,s_q/q_{\rm res})$ & Median error (\%) & Central 95\% (\%) & $N_{10}/100$ & $N_{\rm fail}/3000$ \\

$1\times10^{-4}$ & $(0.25,0.1)$ & $-0.07$ & $[-6.52,7.44]$ & $99$ & $0$\\

$1\times10^{-4}$ & $(0.5,1)$ & $2.47$ & $[-60.07,86.96]$ & $20$ & $0$\\

$1\times10^{-4}$ & $(1,3)$ & $-9.86$ & $[-666.58,412.01]$ & $3$ & $0$\\

$1\times10^{-3}$ & $(0.25,0.1)$ & $-0.01$ & $[-9.39,7.57]$ & $98$ & $0$\\

$1\times10^{-3}$ & $(0.5,1)$ & $-0.65$ & $[-68.32,78.78]$ & $19$ & $0$\\

$1\times10^{-3}$ & $(1,3)$ & $-102.64$ & $[-769.97,670.59]$ & $7$ & $1$\\

\end{tabular}\end{ruledtabular}\end{table}

\begin{figure}[t]\centering

\includegraphics[width=0.82\textwidth]{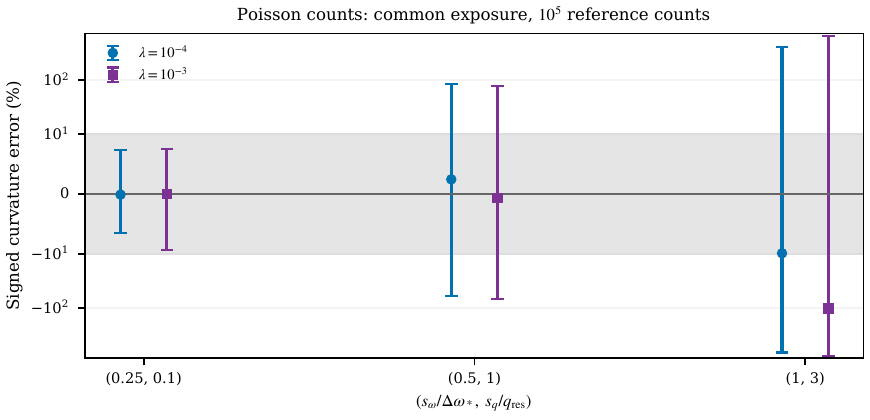}

\caption{Shot-noise effect on reconstructed curvature. Points show medians; bars show central 95\% ranges over 100 repeated measurements. Circles and squares label $\lambda=10^{-4}$ and $10^{-3}$. Each reconstruction uses 30 two-helicity spectra under one exposure, corresponding to about $3\times10^6$ expected total counts. Shading marks $\pm10\%$; a symmetric logarithmic scale shows large errors without removing unsuccessful cases.}\label{fig:SresolutionNoise}\end{figure}

The selected main-figure setting gives 99 and 98 of 100 reconstructions inside the $10\%$ target, with zero optimizer/boundary failures among 3000 spectral fits at each $\lambda$. The frequency and momentum widths are fixed during each reconstruction. Broader-kernel cases quantify the loss of information at the same exposure; this statistical outcome is distinct from convergence of the noiseless numerical quadrature. A finite sample of 100 repetitions does not establish a universal success probability.

Further integration checks use 16384 momentum points at the same refined angular, radial, and frequency sampling. At $(0,1)$ the errors become $0.958\%$ and $2.417\%$, with last changes of $0.744\%$ and $0.854\%$ of the full BdG curvature. The coarse cells remain flagged. Cells at $s_q=3q_{\rm res}$ change their $10\%$ classification under successive 4096-, 8192-, and 16384-point integrations and remain numerically unresolved. They do not define a physical resolution boundary.

A separate refined forward model with 8192 momentum points and 963 frequency bins was used for 100 additional noisy reconstructions per $\lambda$ at $(1,3)$. Its matching five-radius, six-angle noiseless estimates are $F=0.997442$ and $0.970892$. The finite-count central $95\%$ intervals are $[-4.724,7.294]$ and $[-5.177,6.004]$; only 1 and 0 realizations lie within $10\%$ of the full BdG curvature. All optimizers converge. Large statistical uncertainty therefore persists under refinement, although the precise bias remains sensitive. The primary ensembles and this check contain 800 complete reconstructions, comprising 24000 joint two-helicity fits. All estimates and the one primary optimizer failure are retained.

Checks with deliberately unequal polarization totals recover mode areas to $1.07\times10^{-9}$ and dichroism to $8.88\times10^{-10}$. FFT and direct Gaussian convolution agree to $5.55\times10^{-16}$; a prescribed cubic derivative is recovered to $1.55\times10^{-15}$.

\section{Lowest-order decay, density response, and boundary qualifications}
\subsection{Two-quasiparticle threshold and cubic vertices}
The decay analysis in this section refers to the original benchmark completion $J_B=0$. The lowest two-quasiparticle threshold is
\begin{equation}
E_2^{\min}(\mathbf k)=\min_{\mathbf q}
[\omega_1(\mathbf q)+\omega_1(\mathbf k-\mathbf q)].
\end{equation}
All other final branches are no lower at the same momentum, so this is the lowest threshold over final band labels. When the initial frequency is below it, the zero-temperature $1\to2$ channel has no energy-conserving solution~\cite{Beliaev2024}. The cubic Hamiltonian in the rotated fluctuation basis is
\begin{align}
H_3^{(u)}&=2u\rho\sum_\alpha(a_\alpha^{\dagger2}a_\alpha+a_\alpha^\dagger a_\alpha^2),\\
H_3^{(V)}&=V\rho\sum_{\langle\alpha\beta\rangle}
[(a_\alpha+a_\alpha^\dagger)a_\beta^\dagger a_\beta+(a_\beta+a_\beta^\dagger)a_\alpha^\dagger a_\alpha],\\
H_3^{(J)}&=-J\sum_{\tau\in A}(a_1a_2a_3+\mathrm{H.c.}).
\end{align}
Transforming with all Bogoliubov amplitudes gives a vertex symmetric in the two outgoing modes. With ordered final band labels, the probability decay rate is
\begin{equation}
\Gamma_n(\mathbf k)=\frac\pi{|\mathrm{BZ}|}\sum_{m,l}\int d^2q\,
|\mathcal V_{n;ml}|^2\delta[\omega_n(\mathbf k)-\omega_m(\mathbf q)-\omega_l(\mathbf k-\mathbf q)].
\end{equation}
The pole halfwidth at this order is $\Gamma_n/2$. The threshold margins used in the Letter are given in Table~\ref{tab:Scommon}. 
We checked twelve directions and independently re-evaluated the thresholds at $\phi=0,\pi/6$. These finite-direction optimizations are not an analytic all-angle bound. The cubic vertex was checked against a real-space third-order coefficient and exchange symmetry. 

The parent expansion $\omega_{T,L}=cq+\alpha_{T,L}q^3+\cdots$ has $\alpha_T=0.32521027$ and $\alpha_L=0.49306073$. To leading Gaussian order,
\begin{equation}
q_{{\rm dec},n}\simeq
\left[\frac{9a_v}{2c(\alpha_n-\alpha_T/4)}\right]^{1/4}\lambda^{1/4}.
\end{equation}
The numerical onsets along $\hat x$ are $(0.19894165,0.17319741)$ for $\lambda=10^{-4}$ and $(0.34673208,0.29754454)$ for $10^{-3}$. Their Berry scales are 0.12135739 and 0.26145658. The hierarchy $q_{\rm pg}\sim\lambda^{1/2}$, $q_*\sim\lambda^{1/3}$, $q_{\rm dec}\sim\lambda^{1/4}$ belongs to the stated Gaussian expansion; static drag and dynamical quantum corrections constrain extrapolation to arbitrarily small $\lambda$.

\subsection{Other responses and natural boundaries}
The uniform density projections $F_\pm=(F_x\pm iF_y)/\sqrt2$ also select the two circular modes at $\Gamma$, with nonzero weight $W=\rho^2\sqrt{3t_\Sigma/(4a_v+3t_\Sigma)}$. This is distinct from the finite-$q$ tensor current, whose linear vertex vanishes at $q=0$. Density dichroism alone diagnoses chirality, not Chern number.

A continuous representative wall $D(y)=\rho(1,e^{i\varphi(y)},e^{-i\varphi(y)})$ has
\begin{equation}
\varphi(y)=\pi+2\arctan\left[\frac{\tanh[(y-y_0)/(2\ell)]}{\sqrt3}\right],\quad
\ell=\sqrt{t'/t_\Sigma},\quad
\sigma_{\rm DW}=\frac{2\rho^2\sqrt{3t't_\Sigma}}{A_c}
\left(2\sqrt3-\frac{2\pi}{3}\right).
\end{equation}
These follow from the leading fixed-amplitude gradient functional. Lattice optimizations allow all internal complex fields to relax, with three pinned rows at each end. For $\lambda=10^{-3}$ on 768 rows, the fitted width is 31.62404 and tension 0.17318683; for $10^{-4}$ on 2048 rows they are 100.00019 and 0.05478609. The residual gradients are approximately $7.4\times10^{-15}$ and $1.1\times10^{-14}$. The low localized mode overlaps with $\partial_yD$ and is approximately reciprocal, $\omega\simeq c|k_\parallel|$: it is a wall translation mode, not an established chiral Chern channel.

For the bulk define direct, indirect, and projected gaps separately:
\begin{align}
\Delta_{n,n+1}^{\rm dir}&=\min_{\mathbf k}[\omega_{n+1}(\mathbf k)-\omega_n(\mathbf k)],\\
\Delta_{n,n+1}^{\rm ind}&=\min_{\mathbf k}\omega_{n+1}(\mathbf k)-\max_{\mathbf k}\omega_n(\mathbf k),\\
\Delta_{n,n+1}^{\rm proj}(k_\parallel)&=
\min_{k_\perp}\omega_{n+1}(k_\parallel,k_\perp)-\max_{k_\perp}\omega_n(k_\parallel,k_\perp).
\end{align}
At $\lambda=10^{-3}$ the indirect gaps are approximately $-11.139$ and $-8.596$. The direct band isolation therefore does not furnish a clean frequency gap for arbitrary bulk states. Self-consistent free terminations on 64/96 rows show localized states, but some are already present in the conserving parent. No robust isolated unidirectional channel has been established for these terminations. This limitation does not invalidate the bulk band Chern numbers~\cite{Malki2019}; it limits claims about clean edge transport.

\section{Relation to higher-moment phases and geometric probes}
Dipole conservation is part of the higher-moment symmetry framework: restricting motion by charge and dipole conservation, organizing multipole transformations, and gauging spatially dependent phase rotations are established ideas~\cite{Pretko2017Subdimensional,Pretko2018Gauge,Gromov2019Multipole,Gromov2024Review}. At the conserving parent, the neutral dipole condensate preserves ordinary charge symmetry while spontaneously breaking two vector dipole rotations. Finite flavor mixing explicitly breaks these rotations and turns the associated Goldstone modes into pseudo-Goldstone modes. This should be distinguished from charge condensation in fractonic superfluids and from condensation of particles with subdimensional mobility~\cite{Yuan2020,Chen2021FractonicII,Stahl2022Multipole}. Neutral dipolar phases have also been studied in tilted and constrained one-dimensional Bose--Hubbard models~\cite{Lake2023Tilted,Zechmann2023}. Those microscopic examples motivate dipole order; they do not establish a projection to the two-dimensional canonical Hamiltonian used here. Continuum dipole-superfluid theories distinguish symmetry-breaking patterns and their allowed dynamics~\cite{Jain2023,Stahl2023Hydro,Armas2024Ideal}. Their broader hydrodynamic setting does not fix the equal Gaussian stiffnesses or the cubic splitting of this particular lattice model.

Topological excitation bands are likewise established in Bose condensates with imposed gauge structure, pumped chiral backgrounds, and orbital superfluid order~\cite{Furukawa2015,Bardyn2016,Xu2016,DiLiberto2016,Jalali2023}. In particular, excitation-band topology need not be ground-state topological order~\cite{DiLiberto2016}. Our comparison concerns which soft fields acquire the topology and how they are excited: the two fields descend from commuting dipole rotations, and the same weak mixing pins them and supplies the gyroscopic mass. Conventional quadratic-boson diagonalization and paraunitary geometry provide the method~\cite{Colpa1978,Engelhardt2015,Shindou2013,Tesfaye2025}. Positivity implies a real stable spectrum in the strictly positive case~\cite{Peano2018}; the positive lower bound proved here is model-specific and applies at nonzero mixing. It must not be carried unchanged to the zero-mode endpoint.

Berry curvature underlies several distinct measurement routes~\cite{Xiao2010}. Wave-packet dynamics under opposite forces can map curvature~\cite{Price2012}; Bloch-state tomography reconstructs the momentum-dependent state~\cite{Flaschner2016}; integrated excitation rates can reveal quantum-metric components~\cite{Ozawa2018}, and integrated circular dichroism accesses Chern response in an occupied band~\cite{Tran2017,Asteria2019}. Prepared-quasiparticle transitions can access symplectic geometry~\cite{Tesfaye2025}. Our tensor source instead creates one quasiparticle from the condensate vacuum. The normalized mode areas and their radial derivative rely on the specified two-mode matching, common source calibration, and orbital embedding. The cited measurement protocols do not establish that derivative identity or implement this source.

\begin{figure}[t]
\centering
\includegraphics[width=0.96\textwidth]{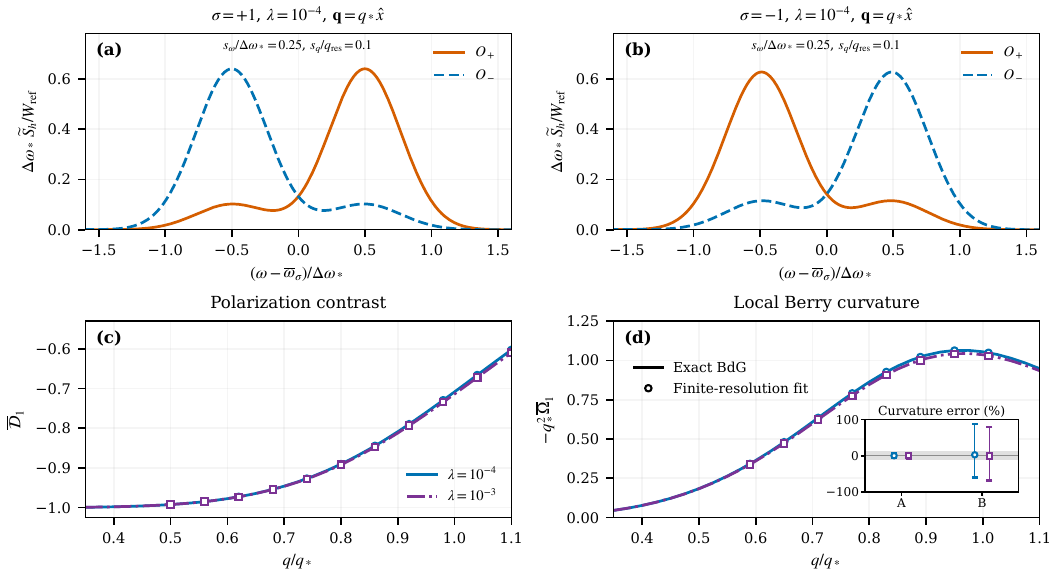}
\caption{Spectral-area extraction and curvature reconstruction. (a,b) Independently calculated opposite-chirality spectra at the same $q_*\hat x$, $\lambda=10^{-4}$, with common calibration. Exact time reversal also reverses momentum. (c) Angular-averaged dichroism $\overline{\mathcal D}_1$ and (d) curvature $\overline{\Omega}_1$: full BdG lines and joint-fit symbols at Gaussian widths $(s_\omega/\Delta\omega_*,s_q/q_{\rm res})=(0.25,0.1)$. Circles denote $\lambda=10^{-4}$ and squares $10^{-3}$. Main curves are noiseless; the inset shows median and central $95\%$ Poisson error ranges at settings A, $(0.25,0.1)$, and B, $(0.5,1)$, with about $3\times10^6$ expected counts per reconstruction. Panel (c) shows the intermediate observable used for the curvature reconstruction in the Letter.}
\label{fig:Scomplete_readout}
\end{figure}

\section{Numerical checks and scope}
Table~\ref{tab:Saudits} summarizes independent numerical checks of the results used in the manuscript. The residuals test implementation consistency within the stated Gaussian description.
\begin{table}[t]
\caption{Selected implementation checks.}\label{tab:Saudits}
\begin{ruledtabular}
\begin{tabular}{lc}
Check&Reported maximum discrepancy\\
Independent bond-list Fourier versus compact blocks&$1.78\times10^{-15}$\\
Real-space finite-difference Hessian versus Bloch frequencies&$3.34\times10^{-9}$\\
Projected $\Gamma$ canonical blocks&$1.78\times10^{-15}$\\
Linear lattice torque Ward identity&$1.46\times10^{-15}$\\
Independent $7\times7$ Peierls-source derivative&$1.79\times10^{-15}$\\
Onsager--Casimir relative residual&$2.16\times10^{-16}$\\
$w=1,2,3$ local-flux quadratures&$4.44\times10^{-16}$
\end{tabular}
\end{ruledtabular}
\end{table}

The present model connects the dipole phase doublet, its topology under weak explicit breaking, and its tensor response. Insulating multipolar phases and topology with dipolar symmetry breaking provide prior dipolar settings~\cite{Fliss2021,MayMann2021,Lam2024,Anakru2024,Zhang2025Tensor}; dipole superfluids and kagome Bogoliubov bands provide the closest condensate settings~\cite{Yuan2020,Lake2022,Jalali2023}. The distinguishing mechanism is the chiral reorganization of two modes that become Goldstone modes at exact conservation. Full-lattice invariants establish their global topology, while polarization spectroscopy probes their local geometry.

A full frequency-dependent one-loop self-energy and the corresponding Ward-consistent one-loop dynamical response remain open. Other extensions concern higher-order and finite-temperature decay, calibration drift and non-Gaussian intrinsic line shapes, a microscopic realization of the model and source, and topology at general boundaries. The present results comprise the Gaussian theory and the explicitly identified corrections.

\makeatletter
\immediate\write\@auxout{\string\citation{ArticleTitleControl}}
\makeatother
\bibliographystyle{apsrev4-2}
\bibliography{references}